\documentclass[aps,prd,twocolumn,groupedaddress]{revtex4}
\usepackage{array,hhline,dcolumn} 
\usepackage{rotating} 
\usepackage{epstopdf}
\usepackage{graphicx}
\usepackage{amssymb}

\begin{document}


\preprint{NuSOnG-PhysicsCase}

\title{Terascale Physics Opportunities at a High Statistics, High Energy \\
Neutrino Scattering Experiment:  NuSOnG
}
\date{\today}

\author{T.~Adams$^5$, P.~Batra$^3$, L.~Bugel$^3$,
  L. Camilleri$^3$, J.M.~Conrad$^3$, A.~de~Gouv\^ea$^{11}$,
  P.H.~Fisher$^8$, J.A.~Formaggio$^8$, J.~Jenkins$^{11}$,
  G.~Karagiorgi$^3$, T.R.~Kobilarcik$^4$, S.~Kopp$^{15}$,
  G.~Kyle$^{10}$, W.A.~Loinaz$^1$, D.A.~Mason$^4$, R.~Milner$^8$, R.~Moore$^4$,
  J.~G.~Morf\'{\i}n$^4$, M.~Nakamura$^9$, D.~Naples$^{12}$,
  P.~Nienaber$^{13}$, F.I~Olness$^{14}$, J.F.~Owens$^5$,
  S.F. Pate$^{10}$, A. Pronin$^{16}$, W.G.~Seligman$^3$,
  M.H.~Shaevitz$^3$, H.~Schellman$^{11}$, I.~Schienbein$^7$,
  M.J.~Syphers$^4$, T.M.P.~Tait$^{2,11}$, T.~Takeuchi$^{16}$,
  C.Y.~Tan$^4$, R.G.~Van~de~Water$^6$,
  R.K.~Yamamoto$^8$, J.Y.~Yu$^{14}$\\
  ~~~\\
}

\affiliation{$^1$\/Amherst College, Amherst, MA 01002 \\
$^2$\/Argonne National Laboratory, Argonne , IL 60439 \\
$^3$\/Columbia University, New York, NY 10027  \\
$^4$\/Fermi National Accelerator Laboratory, Batavia IL 60510 \\
$^5$\/Florida State University, Tallahassee, FL 32306  \\
$^6$\/Los Alamos National Accelerator Laboratory, Los Alamos, NM 87545  \\
$^7$LPSC, Universit\'{e} Joseph Fourier Grenoble 1, 38026 Grenoble, France\\
$^8$\/Massachusetts Institute of Technology, Cambridge, MA 02139  \\
$^9$\/Nagoya University, 464-01, Nagoya, Japan \\
$^{10}$\/New Mexico State University, Las Cruces, NM 88003 \\
$^{11}$\/Northwestern University, Evanston, IL 60208  \\
$^{12}$\/University of Pittsburgh, Pittsburgh, PA 15260  \\
$^{13}$\/Saint Mary's University of Minnesota, Winona, MN 55987\\
$^{14}$\/Southern Methodist University, Dallas, TX 75205 \\
$^{15}$\/University of Texas, Austin TX 78712\\
$^{16}$\/Virginia Tech, Blacksburg VA 24061\\
}

\begin{abstract}
  This article presents the physics case for a new high-energy,
  ultra-high statistics neutrino scattering experiment, NuSOnG
  (Neutrino Scattering on Glass).  This experiment uses a
  Tevatron-based neutrino beam to obtain over an order of magnitude
  higher statistics than presently available for the purely weak
  processes $\nu_{\mu}+e^- \rightarrow \nu_{\mu}+ e^-$ and $\nu_{\mu}+
  e^- \rightarrow \nu_e + \mu^-$.  A sample of Deep Inelastic
  Scattering events which is over two orders of magnitude larger than
  past samples will also be obtained.  As a result, NuSOnG will be
  unique among present and planned experiments for its ability to
  probe neutrino couplings to Beyond the Standard Model physics.  Many
  Beyond Standard Model theories physics predict a rich hierarchy of
  TeV-scale new states that can correct neutrino cross-sections,
  through modifications of $Z\nu\nu$ couplings, tree-level exchanges
  of new particles such as $Z^\prime$s, or through loop-level oblique
  corrections to gauge boson propagators.  These corrections are
  generic in theories of extra dimensions, extended gauge symmetries,
  supersymmetry, and more.  The sensitivity of NuSOnG to this new
  physics extends beyond 5 TeV mass scales.  This article reviews
  these physics opportunities.

\end{abstract}

\pacs{???}
\maketitle

\section{introduction\label{se:intro}}

Exploring for new physics at the ``Terascale'' -- energy scales of
$\sim$ 1 TeV and beyond -- is the highest priority for particle
physics.  A new, high energy, high statistics neutrino scattering
experiment running at the Tevatron at Fermi National Accelerator
Laboratory can look beyond the Standard Model at Terascale energies by
making precision electroweak measurements, direct searches for novel
phenomena, and precision QCD studies.  In this article we limit the
discussion to precision electroweak measurements; QCD studies and
their impact on the precision measurements are explored in ref.
\cite{EOI, QCDPRD}.  The ideas developed in this article were proposed
within the context of an expression of interest for a new neutrino
experiment, NuSOnG (Neutrino Scattering On Glass) \cite{EOI}.

\begin{figure}
{\includegraphics[width=3.in]{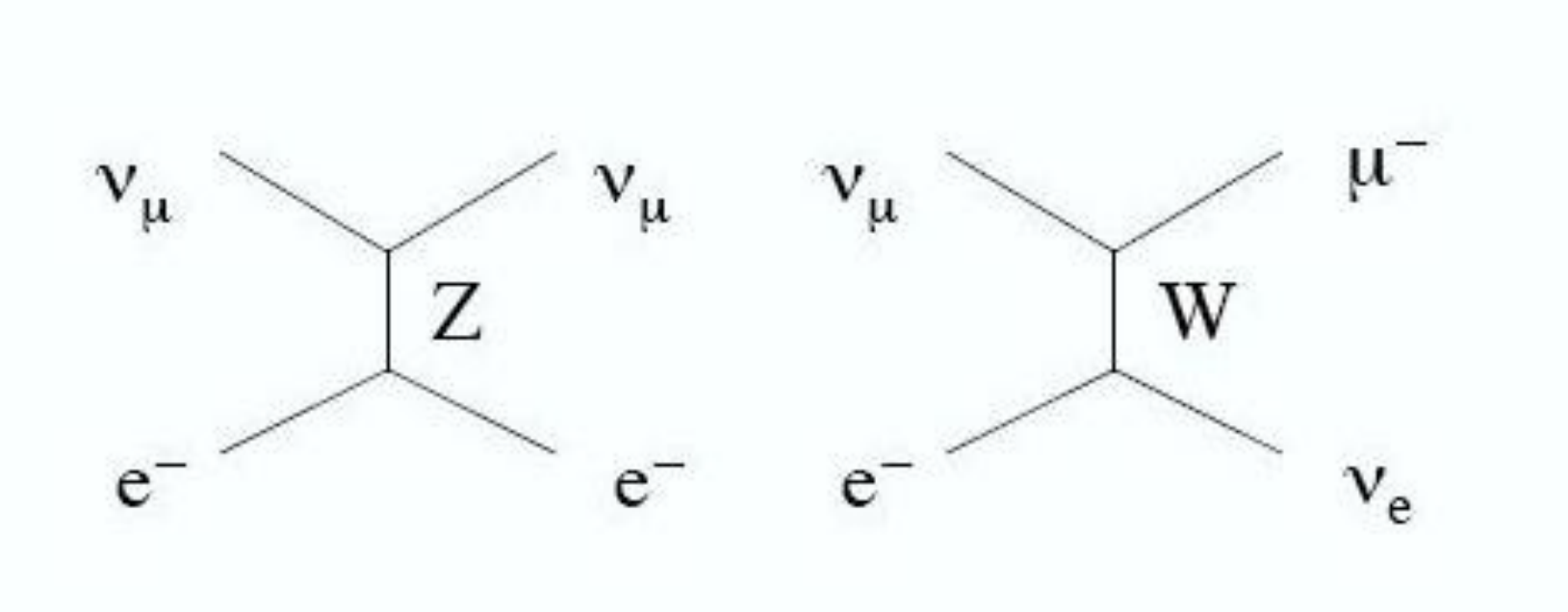}}
\caption{\label{fig:FeynmanESIMD} Left:  ``elastic scattering'' (ES).
Right:  ``Inverse Muon Decay'' (IMD). 
 }
\end{figure}

A unique and important measurement of the NuSOnG physics program is
the ratio of neutral current (NC) and charged current (CC)
neutrino-electron scattering, which probes new physics.  The leading order
Feynman diagrams for these processes are shown in
Fig. \ref{fig:FeynmanESIMD}.  The NC process, $\nu_{\mu}+e^-
\rightarrow \nu_{\mu}+ e^-$, called ``elastic scattering'' or ES,
provides the sensitivity to the Terascale physics.  This process can
explore new physics signatures in the neutrino sector which are not
open to other, presently planned experiments.  The CC process, called
``inverse muon decay'' or IMD, $\nu_{\mu}+ e^- \rightarrow \nu_e +
\mu^-$, is well understood in the Standard Model due to precision
measurement of muon decay \cite{mudk}.  Since the data samples are
collected with the same beam, target and detector at the same time,
the ratio of ES to IMD events cancels many systematic errors while
maintaining a strong sensitivity to the physics of interest.  Our
measurement goal of the ES to IMD ratio is a 0.7\% error, adding
systematic and statistical errors in quadrature.  The high sensitivity
which we propose arises from the combined high energy and high
intensity of the NuSOnG design, leading to event samples more than an
order of magnitude higher than past experiments.

Normalizing the ES to the IMD events represents an important step
forward from past ES measurements, which have normalized neutrino-mode
ES measurements to the antineutrino mode, $\bar \nu_{\mu}+e^- \rightarrow
\bar \nu_{\mu}+ e^-$\cite{ahrens,CHARMIIsin2thw}. The improvement is
in both the experimental and the theoretical aspects of the
measurement.  First, the flux contributing to IMD and $\nu$ ES is
identical, whereas neutrino and antineutrino fluxes are never
identical and so require corrections.  Second, the ratio of $\nu$ ES to
$\bar \nu$ ES cancels sensitivity to Beyond Standard Model (BSM) physics
effects from the NC to CC coupling ratio, $\rho$, which are among the
primary physics goals of the NuSOnG measurement.  In contrast, there
is no such cancellation in the ES to IMD ratio.

\begin{figure}

{\includegraphics[width=3.in, bb=100 300 560 570]{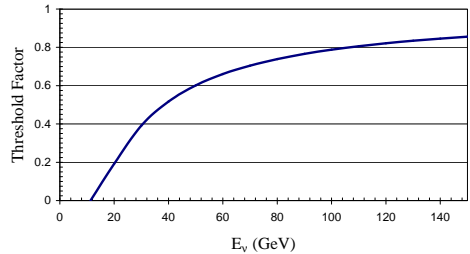}}
\caption{\label{fig:IMDthresh} Threshold factor for the IMD cross section,
as a function of neutrino energy.}
\end{figure}

The design of this experiment, described in
Sec.~\ref{concept}, is driven both by requiring sufficient statistics
to make precision neutrino-electron scattering measurements and by the
need for a neutrino flux which does not extend below the IMD
threshold.  The threshold for IMD events is
\begin{equation}
E_\nu \ge E_\mu \ge {{m_\mu^2}\over{2 m_e}} =10.9~{\rm GeV},
\end{equation}
where we have dropped the small $m_e^2$ term for simplicity.
The functional form above threshold, shown in
Fig.~\ref{fig:IMDthresh}, is given by $(1-m_\mu^2/E_{cm}^2)^2$, where
$E_{cm}$ is the center of mass energy.  Thus a high energy neutrino
beam is required to obtain a high statistics sample of these events.
The flux design should provide a lower limit on the beam energy of
about 30 GeV, still well above the IMD threshold.

Sec.~\ref{EWreview} describes the Standard Model Physics of
neutrino electroweak scattering, for both electron and quark targets.
In this section, the value of the normalization of the ES to IMD
events is further explored.  The very high statistics will also permit
an electroweak measurement using the deep inelastic scattering (DIS)
data sample from NuSOnG, via the ``Paschos Wolfenstein method'' (PW)
\cite{PW}.  The best electroweak measurement using DIS events to date
comes from the NuTeV experiment, which has observed an anomaly.  The
status of this result is reviewed below.  Making conservative
assumptions concerning systematic improvements over NuTeV, our
measurement goal using this technique is a 0.4\% error on
$\sin^2\theta_W$, adding statistical and systematic errors in
quadrature.

In Sec.~\ref{Terascale}, we discuss NuSOnG's potential to discover or
constrain new physics through indirect probes, by making precision
measurements of SM processes to look for deviations from SM
predictions.  We first frame the issue by considering in turn several
model-independent parameterizations of possible new physics and asking
what constraints will be imposed on new physics in the event NuSOnG
agrees with the SM.  (1) Oblique correction parameters describe the
effects of heavy new states in vector boson loops.  (2) New states may
induce higher-dimensional effective operators involving neutrinos.
Finally, (3) new states may modify the couplings of the gauge bosons
to neutrinos and leptons, including possibly violating lepton
universality.  In each case we consider the ability of NuSOnG to
detect or constrain these types of deviations from the SM.

In Sec.~\ref{Models}, we examine specific models for new physics.  We
begin by presenting the sensitivity to a set of new physics models.
In particular, we consider
\begin{itemize}
\item typical $Z^\prime$ models,
\item non-degenerate leptoquark models,
\item R-parity violating SUSY models,
\item extended Higgs models.
\end{itemize} 
The models were selected because they are often used as benchmarks in
the literature.  While this list is not exhaustive, it serves to
illustrate the possibilities.  For each case, we consider how NuSOnG
compares to other measurements and note the unique contributions.  We
end this section by approaching the question from the opposite view,
asking: how could the results from NuSOnG clarify the underlying
physics model, should evidence of new physics emerge from LHC in the
near future?

Two further studies which can be performed by NuSOnG are QCD
measurements and direct searches.  The very large ($\sim 600$ million
event) DIS sample will allow the opportunity for precision studies of
QCD.  There are many interesting measurements which can be made in
their own right and which are important to NuSOnG's Terascale physics
program. The very high flux will also permits direct searches for new
physics.  Those which complement the physics discussed in this paper
include:
\begin{itemize}
\item non-unitarity in the light neutrino mixing matrix;  
\item wrong-sign inverse muon decay (WSIMD),
$\bar{\nu}_\mu + e^- \rightarrow \mu^- + \bar{\nu}_e$;
\item decays of 
neutrissimos, {\it i.e.,} moderately-heavy neutral-heavy-leptons, with
masses above 45 GeV.
\end{itemize}
For more information on these studies, see refs.~\cite{EOI, QCDPRD}.

\section{Conceptual Design for the Experiment \label{concept}}

In order to discuss the physics case for a new high energy, high
statistics experiment, one must specify certain design parameters for
the beam and detector.  The beam and detector should marry the best
aspects of NuTeV \cite{NuTeVbeam}, the highest energy neutrino
experiment, and Charm II \cite{CharmIIdet}, the experiment 
with the largest ES sample to date.  The plan presented here is not
optimized, but provides a basis for discussion.  The final design of the
NuSOnG detector will be based on these concepts, and is still 
under development.

In this section, we present, but do not justify, the design choices.
Later in this article, we discuss the reasoning for the choices,
particularly in Secs.~\ref{IMDvsBarNu} and \ref{justify}.

\begin{figure}
{\includegraphics[width=3.5in, bb=0 400 560 670]{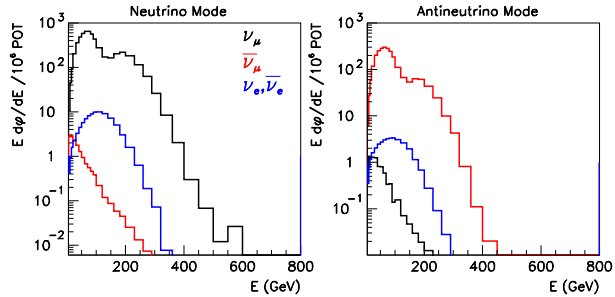}}
\caption{The assumed energy-weighted flux, from the NuTeV Experiment
  \cite{NuTeVbeam}, in neutrino mode (left) and antineutrino mode
  (right).  Black: muon neutrino, red: muon antineutrino,
  blue: electron neutrino and antineutrino flux.}
\label{fig:beam}
\end{figure}

We will assume a beam design based on the one used by the NuTeV experiment
\cite{NuTeVbeam}, which is the most recent high energy neutrino
experiment.  This experiment used 800 GeV protons on target.  The beam
flux, shown in Fig.~\ref{fig:beam}, is ideal for the physics case
for several reasons.  There is essentially no flux below 30 GeV,
hence all neutrinos are well above the IMD threshold. It is sign-selected:
in neutrino mode, 98.2\% of neutrino interactions were due to $\pi^+$
and $K^+$ secondaries, while in antineutrino mode 97.3\% came from
$\pi^-$ and $K^-$.  The ``wrong sign'' content was very low, with a
0.03\% antineutrino contamination in neutrino mode and 0.4\% neutrino
contamination in antineutrino mode.  The electron-flavor content was
1.8\% in neutrino mode and 2.3\% in antineutrino mode.  The major
source of these neutrinos is $K^{\pm}_{e3}$ decay, representing 1.7\%
of the total flux in neutrino mode, and 1.6\% in antineutrino mode.

Redesign of the beamline for NuSOnG is expected to lead to modest 
changes in these ratios.  For example, if the decay pipe length is
1.5 km rather than 440 m, as in NuTeV, the $\pi/K$ ratio increases by 20\%
and the fractional $\nu_e$ content is reduced.

With respect to Tevatron running conditions, we will assume that
twenty times more protons on target (POT) per year can be produced for
NuSOnG compared to NuTeV.  This is achieved through three times higher
intensity per pulse (or ``ping''). Nearly an order of magnitude
more pulses per spill are provided.  Our studies assume 4 $\times$ 10$^{19}$
POT/year, with 5 years of running.    Preliminary studies supporting
these goals are provided in ref. \cite{pot}.

The event rates quoted below are consistent with 1.5$\times10^{20}$
protons on target in neutrino running and $0.5\times10^{20}$ protons
on target in antineutrino running.  The choice to emphasize neutrino
running is driven by obtaining high statistics ES, which has a higher
cross section for neutrino scatters, and to use the IMD for
normalization -- this process only occurs in neutrino scattering.  The
Standard Model forbids an IMD signal in antineutrino mode.  However,
some antineutrino running is required for the physics described in the
following sections, especially the PW electroweak measurement.

The beam from such a design is highly forward directed.  NuTeV was
designed so that 90\% of the neutrinos from pion decay were contained
within the detector face, where the detector was located at 1 km.  For
NuSOnG, which will use a 5 m detector, $\sim$90\% of the neutrinos from pion
decay are contained at $\sim$3 km.

The optimal detector is a fine-grained calorimeter for electromagnetic
shower reconstruction followed by a toroid muon spectrometer. This
allows excellent reconstruction of the energy of the outgoing lepton
from charged current events. We employ a Charm II style design
\cite{CharmIIdet}, which uses a glass target calorimeter followed by a
toroid.  We assume one inch glass panels with active detectors
interspersed for energy and position measurement. Glass provides an
optimal choice of density, low enough to allow electromagnetic showers
to be well sampled, but high enough that the detector length does not
compromise acceptance for large angle muons by the toroid.  Approximately
10\% of the glass will be doped with scintillator to allow
for background studies, as discussed in Sec.~\ref{justify}.

The design introduces four identical sub-detectors of this glass-calorimeter 
and toroid design,
each a total of 29 m in length (including the toroid).  Between
each sub-detector is a 15 m decay
region for direct searches for new physics.
The total fiducial volume is 3 ktons.  

The NuSOnG run plan, for reasons discussed in Sec.~\ref{IMDsub} and
\ref{IMDvsBarNu}, concentrates on running in neutrino mode.  This
design will yield the rates shown in Table~\ref{Tab:rates}.  These
rates, before cuts, are assumed throughout the rest of the discussion.
We can compare this sample to past experiments.  The present highest
statistics sample for $\nu_\mu$ and $\bar \nu_\mu$ ES is from
CHARM~II, with 2677$\pm$82 events in neutrino mode and 2752$\pm$88
events in antineutrino mode \cite{CHARMIIsin2thw}.  Thus the proposed
experiment will have a factor of 30 (2.5) more $\nu$($\bar
\nu$)-electron events.  As an example, after cuts, the first method of
analysis described in Sec.~\ref{justify} retains 63\% of the $\nu$
sample. For deep inelastic scattering, 600M and 190M events are
expected in neutrino and antineutrino modes, respectively.  After
minimal cuts to isolate DIS events \cite{SamThesis},
NuTeV had 1.62M DIS (NC+CC) events in neutrino mode and 0.35M in
antineutrino mode; thus NuSOnG has orders of magnitude more events.

\begin{table}
\begin{ruledtabular}
\begin{tabular}{c|c}
600\/M & $\nu_\mu$ CC Deep Inelastic Scattering\\
190\/M & $\nu_\mu$ NC Deep Inelastic Scattering \\
75\/k & $\nu_\mu$ electron NC elastic scatters (ES) \\
700\/k &$\nu_\mu$ electron CC quasi-elastic scatters (IMD) \\
33\/M &  $\bar \nu_\mu$ CC Deep Inelastic Scattering \\
12\/M & $\bar \nu_\mu$ NC Deep Inelastic Scattering \\
7\/k & $\bar \nu_\mu$ electron NC elastic~scatters (ES)\\
0\/k &  $\bar \nu_\mu$ electron CC quasi-elastic scatters (WSIMD)\\
\end{tabular}
\end{ruledtabular}
\caption{Rates assumed for this paper.   NC indicates ``neutral current''
and CC indicates ``charged current.''  \label{Tab:rates}
}
\end{table}

The detector will incorporate several specialized regions.  A region
of fine vertex-tracking facilitates measurements of the strange
sea relevant for the electroweak analysis, as described in
ref.~\cite{QCDPRD}.  Two possibilities are under consideration: an
emulsion detector or a silicon detector of the style of NOMAD-STAR
\cite{NOMAD-STAR}.  Both are compact and easily accommodated.  For
further QCD studies, it will also be useful to intersperse alternative
target materials: C, Al, Fe, and Pb \cite{QCDPRD}.

\section{Electroweak Measurements in Neutrino Scattering\label{EWreview}}


Neutrino neutral current (NC) scattering is an ideal probe for new
physics.  An experiment like NuSOnG is unique in its ability to test
the NC couplings by studying scattering of neutrinos from both
electrons and quarks. A deviation from the Standard Model predictions
in both the electron and quark measurements would present a compelling
case for new physics.

The exchange of the $Z$ boson between the neutrino $\nu$ and fermion $f$
leads to the effective interaction:
\begin{eqnarray}
\mathcal{L}
& = & -\sqrt{2}G_F
\Bigl[\, \bar{\nu}\gamma_\mu\bigl(g_V^\nu - g_A^\nu \gamma_5\bigr)\nu \,\Bigr]
\Bigl[\, \bar{f}\gamma^\mu\bigl(g_V^f - g_A^f \gamma_5\bigr)f \,\Bigr] \cr
& = & -\sqrt{2}G_F
\Bigl[\, g_L^\nu\,\bar{\nu}\gamma_\mu(1-\gamma_5)\nu
     + g_R^\nu\,\bar{\nu}\gamma_\mu(1+\gamma_5)\nu \,\Bigr] \cr
& & \qquad\qquad
\times
\Bigl[\, g_L^f \,\bar{f}\gamma^\mu(1-\gamma_5)f
     + g_R^f \,\bar{f}\gamma^\mu(1+\gamma_5)f \,\Bigr] \;, \cr
& & 
\label{eq:geneffint}
\end{eqnarray}
where the Standard Model values of the couplings are:
\begin{eqnarray}
g_L^\nu & = & \sqrt{\rho}\left(+\frac{1}{2}\right) \;,\cr
g_R^\nu & = & 0\;, \cr
g_L^f & = & \sqrt{\rho}\left(I_3^f - Q^f\sin^2\theta_W \right) \;,\cr
g_R^f & = & \sqrt{\rho}\left(-Q^f\sin^2\theta_W\right) \;,
\end{eqnarray}
or equivalently,
\begin{eqnarray}
g_V^\nu \;=\; g_L^\nu + g_R^\nu & = & \sqrt{\rho}\left(+\frac{1}{2}\right)\;,\cr
g_A^\nu \;=\; g_L^\nu - g_R^\nu & = & \sqrt{\rho}\left(+\frac{1}{2}\right)\;,\cr
g_V^f   \;=\; g_L^f + g_R^f & = & \sqrt{\rho}\left(I_3^f - 2Q^f\sin^2\theta_W\right) \;,\cr
g_A^f   \;=\; g_L^f - g_R^f & = & \sqrt{\rho}\left(I_3^f\right) \;.
\end{eqnarray}
Here, $I_3^f$ and $Q^f$ are the weak isospin and electromagnetic
charge of fermion $f$, respectively.  In these formulas, $\rho$ is the
relative coupling strength of the neutral to charged current
interactions ($\rho=1$ at tree level in the Standard Model). The weak
mixing parameter, $\sin^2 \theta_W$, is related (at tree level) to
$G_F$, $M_Z$ and $\alpha$ by
\begin{equation}
\sin^2 2 \theta_W=\frac{4 \pi \alpha  }{ \sqrt{2} G_F M_Z^2} .
\end{equation}

\subsection{Neutrino Electron Elastic Scattering \label{nuesub}
}

The differential cross section for $\nu_\mu$ and $\bar \nu_\mu$
ES, defined using the coupling constants
described above, is:
\begin{eqnarray}
\frac{d\sigma}{dT} & = &
\frac{2G_F^2 m_e}{\pi}
\Biggl[ 
(g_L^\nu g_V^e \pm g_L^\nu g_A^e)^2 
\Biggr.\cr
& & \qquad\qquad + 
(g_L^\nu g_V^e \mp g_L^\nu g_A^e)^2 
\left(1-\frac{T}{E_\nu}\right)^2 \cr
& & \qquad\qquad - 
\Biggl. 
\Bigl\{ (g_L^\nu g_V^e)^2   - (g_L^\nu g_A^e  )^2 \Bigr\}
\frac{m_e T}{E_\nu^2}
\Biggr] \;.
\end{eqnarray}
The upper and lower signs correspond to the neutrino and
anti-neutrino cases, respectively.  In this equation, $E_{\nu}$ is the
incident ${\nu}_{\mu}$ energy and $T$ is the electron recoil kinetic
energy.

More often in the literature, the cross section is defined in terms of
the parameters $(g_{V}^{\nu e},g_{A}^{\nu e})$, which are defined as
\begin{eqnarray}
g_V^{\nu e} 
& \equiv & (2g_L^\nu g_V^e)
\;=\; \rho\left(-\frac{1}{2}+2\sin^2\theta_W\right) \;,\cr
g_A^{\nu e}
& \equiv & (2g_L^\nu g_A^e)
\;=\; \rho\left(-\frac{1}{2}\right) \;,
\end{eqnarray}
In terms of these parameters, we can write:
\begin{eqnarray}
\frac{d\sigma}{dT} & = &
\frac{G_F^2 m_e}{2\pi}
\Biggl[ 
(g_V^{\nu e} \pm g_A^{\nu e})^2 
\Biggr. \cr
& & + 
(g_V^{\nu e} \mp g_A^{\nu e})^2 
\left(1-\frac{T}{E_\nu}\right)^2 \cr
& &  - 
\Biggl. 
\Bigl\{ (g_V^{\nu e})^2   - (g_A^{\nu e}  )^2 \Bigr\}
\frac{m_e T}{E_\nu^2}
\Biggr] \;.
\end{eqnarray}
When $m_e \ll E_\nu$, as is the case in NuSOnG, the third term in
these expressions can be neglected.  If we introduce the variable
$y=T/E_\nu$, then
\begin{eqnarray}
\frac{d\sigma}{dy}
& = & \frac{G_{F}^{2}m_e E_\nu}{2\pi}
\left[ 
  \left( g_{V}^{\nu e} \pm g_{A}^{\nu e} \right)^{2}
+ \left( g_{V}^{\nu e} \mp g_{A}^{\nu e} \right)^{2}
  \left( 1 - y \right)^{2} 
\right]\;.\cr
& &
\label{differentialcrosssection}
\end{eqnarray}
Integrating, we obtain the total cross sections
which are
\begin{eqnarray}
\sigma & = & 
\frac{G_{F}^{2}m_e E_\nu}{2\pi}
\left[ 
  \left( g_{V}^{\nu e} \pm g_{A}^{\nu e} \right)^{2}
+ \frac{1}{3}\left( g_{V}^{\nu e} \mp g_{A}^{\nu e} \right)^{2}
\right]\;.
\label{sigma_enu}
\end{eqnarray}
Note that
\begin{eqnarray}
\left( g_{V}^{\nu e} + g_{A}^{\nu e} \right)^{2}
& =&  \rho^2\left(-1+2\sin^2\theta_W\right)^2  
\; \cr
& = &\; \rho^2\left(1-4\sin^2\theta_W+4\sin^4\theta_W\right) \;,\cr
\left( g_{V}^{\nu e} - g_{A}^{\nu e} \right)^{2}
& = & \rho^2\left(2\sin^2\theta_W\right)^2
\; \cr
& =& \; \rho^2\left(4\sin^4\theta_W\right) \;.
\end{eqnarray}
Therefore,
\begin{eqnarray}
\sigma(\nu_{\mu}\, e) & = & \frac{G_F^2 m_e E_\nu}{2\pi}
\,\rho^2
\Biggl[ 1 - 4\sin^2\theta_W + \frac{16}{3}\sin^4\theta_W \Biggr] \;,\cr
\sigma({\bar \nu_{\mu}}\, e) & = & \frac{G_F^2 m_e E_\nu}{2\pi}
\,\frac{\rho^2}{3}
\Biggl[ 1 - 4\sin^2\theta_W + 16\sin^4\theta_W \Biggr] \;. \cr
& & 
\end{eqnarray}

\begin{figure}
\vspace{-0.75in}
\scalebox{0.45}{\includegraphics{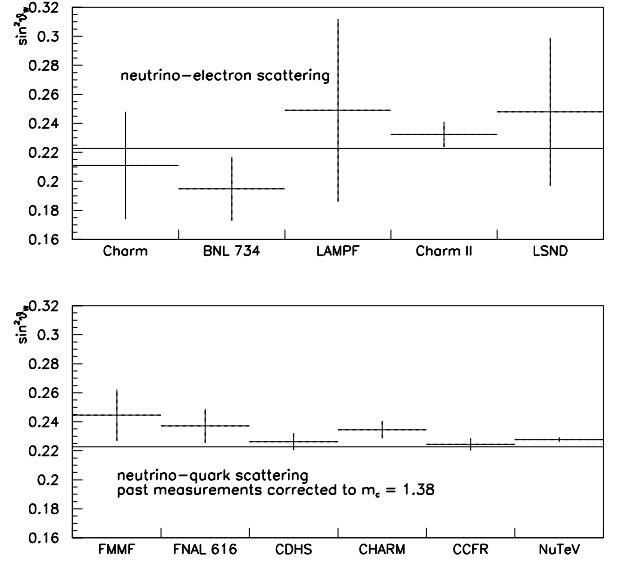}}
\vspace{-0.75in}
\caption{\label{pastnu} Measurements of $\sin^2 \theta_W$ from past
  experiments.  Top: neutrino-electron elastic scattering experiments.
  Bottom: neutrino DIS experiments.  All DIS results are adjusted to
  the same charm mass (relevant for experiments not using the PW method).
  The Standard Model value, indicated by the line, is $0.2227$ \cite{PDG}.}
\end{figure}

The ratio of the integrated cross sections for neutrino to antineutrino
electron ES is
\begin{equation}
R_{\nu/\bar \nu}
\;=\; \frac{\sigma(\nu_{\mu}\, e)}{\sigma^({\bar \nu_{\mu}} e)}
\;=\; 3\;
\frac{1-4\sin^2\theta_W+{{16}\over{3}}\sin^4\theta_W}
      {1-4\sin^2\theta_W+16\sin^4\theta_W} \;.  \label{eq:nuerat}
\end{equation}
Fig. \ref{pastnu}(top) shows the results for $\sin^2 \theta_W$ 
from many past experiments which have used this ``$\nu/\bar \nu$ ES
ratio.''

In the ratio, $R_{\nu/\bar \nu}$, the dependence on $\rho$ canceled.
This directly extracts $\sin^2\theta_W$.     The relationship
between the error on the ratio and the error on $\sin^2\theta_W$,
which for convenience we abbreviate as $z$, is:
\begin{eqnarray}
\delta z  & = &( \frac{32z-12}{16z^{2}-4z+1} + \nonumber \\
& ~~& \frac{448z^{2}-144z-512z^{3}+12}{48z^{2}-8z-128z^{3}+256z^{4}+1}) ^{-1}\delta R_{\nu/\bar \nu} \nonumber\\
& = & -0.103\;\delta R_{\nu/\bar \nu};\\
\delta z/z  & = & -0.575\;\delta R_{\nu/\bar \nu}/R_{\nu/\bar \nu},
\label{rationunubarerr}\end{eqnarray}
for $z=0.2227$ (or $R_{\nu/\bar \nu}=1.242$).
Roughly, the fractional error on $\sin^2\theta_W$ is 60\% of the 
fractional error on $R_{\nu/\bar \nu}$.

\subsection{A New Technique: Normalization Through IMD \label{IMDsub}
}

\begin{figure*}
\scalebox{0.425}{\includegraphics{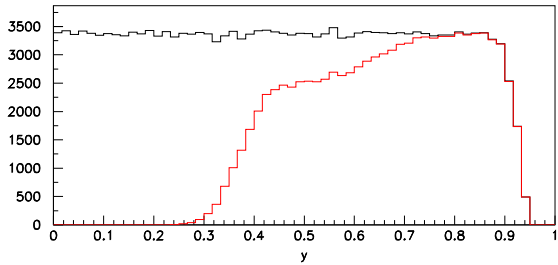}}~\scalebox{0.425}{\includegraphics{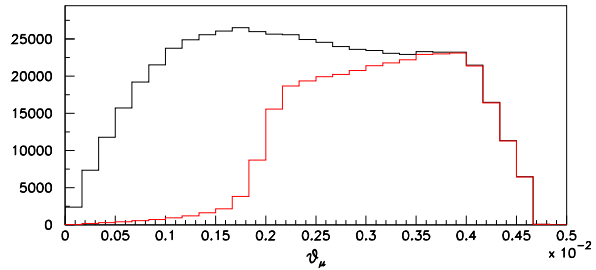}}
\vspace{-2.5in}
\caption{\label{fig:ycut} Kinematic distributions for IMD events from incident neutrino energy between 100 and 200 GeV.  Left: $y$ distribution;  right: 
$\theta_\mu$ distribution.  Black: distribution of events before cuts; Red:
distribution after cuts for analysis method 1 (see Sec.~\ref{justify}).}
\end{figure*}

An experiment such as 
NuSOnG can make independent measurements of the electroweak
parameters for both $\nu_\mu$ and $\bar \nu_\mu$-electron scattering.
We can achieve this via ratios or by direct extraction of the cross
section.    In the case of $\nu_\mu$-electron scattering, we
will use the ratio of the number of events in 
neutrino-electron elastic scattering to inverse muon decay:
\begin{equation} 
{{N(\nu_\mu e^- \rightarrow \nu_\mu e^-)}
\over{N(\nu_\mu e^- \rightarrow \mu^- \nu_e)}}
=
\frac{\sigma^{\nu e}_{NC} \times \Phi^\nu}{\sigma^{IMD} \times \Phi^{\nu}}.
\end{equation}
Because the cross section for IMD events is well determined by the
Standard Model, this ratio should have low errors and will isolate the
EW parameters from NC scattering.  In the discussion below, 
we will assume that the systematic error on this ratio is 0.5\%.

In the case of $\bar \nu_\mu$ data, the absolute normalization is more
complex because there is no equivalent process to inverse muon decay
(since there are no positrons in the detector).  One can use the fact
that, for low exchange energy (or ``nu'') in Deep Inelastic
Scattering, the cross sections in neutrino and antineutrino scattering
approach the same constant, $A$ \cite{lownu}.  This is called the ``low nu
method'' of flux extractions.  For DIS events with low
energy transfer and hence low hadronic energy ($5 \lesssim E_{had}
\lesssim 10$ GeV), $N^{low~E_{had}}_{\nu DIS} = \Phi^\nu A$ and
$N^{low~E_had}_{\bar\nu DIS} = \Phi^{\bar \nu} A$.  The result is that
the electroweak parameters can be extracted using the ratio
\begin{equation}
\frac {{N^{low~E_{had}}_{\nu DIS}}} {N^{low~E_{had}}_{\bar\nu DIS}}  \times
\frac {N(\bar \nu_\mu e^- \rightarrow \bar \nu_\mu e^-)}
{N(\nu_\mu e^- \rightarrow \mu^- \nu_e)}
=
\frac {\Phi^\nu} {\Phi^{\bar \nu}} \times
\frac {\sigma^{\bar \nu e}_{NC} \times \Phi^{\bar \nu}}
{\sigma^{IMD} \times \Phi^\nu}.
\end{equation}
The first ratio cancels the DIS cross section, leaving the
energy-integrated $\nu$ to $\bar \nu$ flux ratio.  The IMD events in
the denominator of the second term cancel the integrated $\nu$ flux.
The NC elastic events cancel the integrated $\bar \nu$ flux.

Because of the added layer of complexity, the antineutrino ES
measurement would have a higher systematic error than the neutrino
ES scattering measurement.  The potentially higher error is one factor
leading to the plan that NuSOnG
concentrate on neutrino running for the ES studies.

\begin{figure}
\vspace{-0.5in}
\scalebox{0.425}{\includegraphics{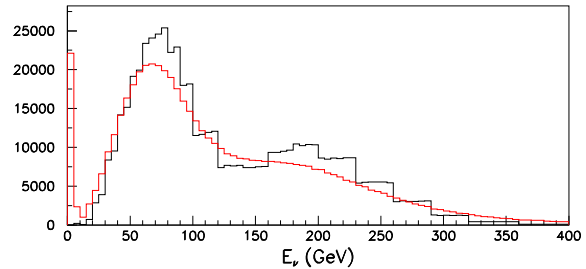}}\vspace{-2.25in}
\caption{\label{fig:EnuIMD} Reconstructed neutrino energy (red) for IMD
events before cuts compared to true neutrino energy (black).}
\end{figure}

As shown in Fig.~\ref{fig:IMDthresh}, IMD events have a kinematic threshold
at 10.9 GeV.   These events also have other interesting kinematic properties.
The minimum energy of the outgoing muon in the lab frame is given by
\begin{equation}
E_{\mu~lab}^{min} = \frac{m_\mu^2 + m_e^2}{2 m_e}= 10.9~{\rm GeV}.
\end{equation}
In the detector described above, muons of this energy and higher will
reach the toroid spectrometer without ranging-out in the glass.  An
interesting consequence is that, independent of $E_\nu$, the energy
transfer in the interaction has a maximum value of
\begin{equation}
y_{max} = 1 - \frac{10.9~{\rm GeV}}{E_\nu}.
\end{equation}
Thus at low $E_\nu$, the cutoff in $y$ is less than unity,
as shown in Fig.~\ref{fig:ycut} (left).  The direct consequence of
this is a strong cutoff in angle of the outgoing muon, shown in
Fig.~\ref{fig:ycut} (right).  In principle, one can reconstruct 
the full neutrino energy in these events:
\begin{equation}
E_{\nu}^{IMD}=\frac{1}{2}\frac{2m_{e}E_{\mu}-m_{e}^{2}-m_{\mu}^{2}}%
{m_{e}-E_{\mu}+p_{\mu}\cos\theta_{\mu}}
\end{equation}
This formula depends on $\theta_\mu$, which is small. The reconstructed
$E_\nu$ is smeared by resolution effects as seen in
Fig.~\ref{fig:EnuIMD}.  While the analysis can be done by summing over
all energies, these distributions indicate that an energy binned
analysis may be possible.  This is more powerful because one can fit
for the energy dependence of backgrounds.  For the illustrative
analyzes below, however, we do not employ this technique.

The error on $\sin^2\theta_W$ extracted from this ratio, $R_{ES/IMD}$,
assuming a Standard Model value for $\rho$, is the same as the error
on the ratio:
\begin{equation}
{{\delta(sin^2\theta_W)}\over{sin^2\theta_W}} \approx {{\delta R_{ES/IMD}}\over {R_{ES/IMD}}}.
\label{ratioesimderr}
\end{equation}

Ref.~\cite{MarcianoParsa} provides a useful summary of radiative corrections 
for the ES and IMD processes, which were originally calculated 
in Ref.~\cite{SSM}.  The error from radiative corrections is expected to
be below 0.1\%.   It is noted that to reduce the error below 0.1\%,
leading two-loop effects must be included.   A new evaluation of the 
radiative corrections is underway \cite{newradcor}.

\subsection{IMD Normalization vs. $\bar \nu$ Normalization
\label{IMDvsBarNu}
}

NuSOnG can measure both the $\nu/\bar \nu$ ES ratio, as in the 
case of past experiments shown in
Eq.~(\ref{eq:nuerat}), as well as the ES/IMD ratio.  In the case of the
former, to obtain the best measurement in a 5 year run, one would
choose a 1:3 ratio of run time in $\nu$ versus $\bar \nu$ mode.  In the
latter case, one would maximize running in $\nu$ mode.  The result of
the two cases is a nearly equal error on $\sin^2\theta_W$, despite
the fact that the error on the $\nu/\bar \nu$ ES is nearly twice that
of the ES/IMD ratio.  To understand this, compare
Eq.~(\ref{rationunubarerr}) to Eq.~(\ref{ratioesimderr}).  However, the
ES/IMD ratio is substantially stronger for reasons of physics.
Therefore, our conceptual design calls for running mainly with a $\nu$
beam.  In this section we explore the issues for these two methods of
measurement further.  We also justify why the precision measurement
requires high energies, only available from a Tevatron-based beam.

\subsubsection{Comparison of the Two Measurement Options}

From the point of view of physics, The ES/IMD ratio is more
interesting than the $\nu/\bar\nu$ ES ratio.  This is because $\rho$
has canceled in the $\nu/\bar \nu$ ES ratio of Eq.~(\ref{eq:nuerat}),
leaving the ratio insensitive to physics which manifests itself
through changes in the NC coupling.  Many of the unique physics goals
of NuSOnG, discussed in Sec.~\ref{Terascale}, depend upon sensitivity
to the NC coupling.

An equally important concern was one of systematics.  The $\nu$ and $\bar \nu$
fluxes for a conventional neutrino beam are substantially different.
For the case of NuSOnG, the fluxes are compared in
Fig.~\ref{fig:beam}.  Predicting the differences in these fluxes
from secondary production measurements and simulations leads to
substantial systematic errors.  For beams at high energies ($>30$
GeV), such as NuSOnG, the ``low nu'' method \cite{lownu} for
determining the ratio of the neutrino to antineutrino fluxes from Deep
Inelastic events, developed by CCFR and NuTeV and described in 
Sec.~\ref{IMDsub}, can be employed.
However, this leads to the criticism that one has introduced a new
process into the purely-leptonic analysis.

Neither criticism is relevant to the ES/IMD ratio. The sensitivity
to the new physics through the couplings does not cancel.  Because
both processes are in neutrino mode, the flux exactly cancels, as long
as the neutrino energies are well above the IMD threshold (this will
be illustrated in the analysis presented in Sec.~\ref{justify}).  This
ratio has the added advantage of needing only neutrino-mode running,
which means that very high statistics can be obtained.  This is
clearly the more elegant solution.

It should be noted that nothing precludes continued running of NuSOnG
beyond the 5-year plan presented here.    This run-length was selected 
as ``reasonable'' for first results.    If interesting physics 
is observed in this first phase, an extended run in antineutrino mode
may be warranted, in which case {\it both} the ES/IMD and $\nu/\bar \nu$ ES
ratios could be measured.  The latter would then constrain $\sin^2\theta_W$
in a pure neutrino measurement and the former is then used to 
extract $\rho$.

To measure the ES/IMD ratio to high precision, there must be little
low energy flux.  This is because the IMD has a threshold of 10.9 GeV,
and does not have substantial rate until $\sim 30$ GeV.
The low-energy cut-off in the flux (see Fig.~\ref{fig:beam})
coming from the energy-angle correlation of neutrinos from pion
decay, is ideal.

\subsubsection{Why a Tevatron-based Beam is Best for Both Options}

The ES/IMD measurement is not an option for the planned beams from the
Main Injector at Fermilab.  For both presently planned Main Injector
experiments at Fermilab \cite{Minerva} and for the proposed 
Project-X DUSEL beam \cite{PXDUSEL}, the
neutrino flux is peaked at $\sim$5 GeV.  The majority of the flux of
these beams is below 5 GeV, and most of the flux is below the 10.9 GeV
IMD threshold.  Because of this, one simply cannot use the IMD events
to normalize.

In principle, the $\nu/\bar \nu$ ES ratio could be used.  However, in
practice this will have large systematics.  The $\nu$ and $\bar \nu$
fluxes for a horn beam are significantly different.  First principles
predictions of secondary mesons are not sufficient to reduce this
error to the precision level.  The energy range is well below the deep
inelastic region where the ``low nu'' method can be applied to
accurately extract a $\bar \nu/\nu$ flux ratio.  Other processes, such
as charged-current quasi-elastic scattering, could be considered for
normalization, but the differences in nuclear effects in neutrino and
antineutrino scattering for these events is not sufficiently well
understood to yield a precision measurement.

Lastly, the ES rates for the present Main Injector beams are too low
for a high statistics measurement.  This is because the cross section
falls linearly with energy. Event samples on the order of 10k may be
possible with extended running in the Project X DUSEL beam in the
future.  From the point of view of statistics, even though two orders
of magnitude more protons on target are supplied in such a beam, the
Tevatron provides a substantially higher rate of ES per year of
running.

Compared to the Main Injector beam, a Tevatron-based beam does not face
these issues.  The choice of running in neutrino mode provides the
highest precision measurement while optimizing the physics.

\begin{table*}[tbp] \centering
\begin{ruledtabular}
\begin{tabular}
[c]{ll|cc|l}\hline
\multicolumn{2}{l|}{Quantity} & Assumed Value & Uncertainty & Source of
Estimate\\\hline
\multicolumn{2}{l|}{Muon} &  &  & \\
& Energy Resolution & $\delta E/E=10\%$ & 2.5\% & NuTeV testbeam measurement\\
& Energy Scale Error & $E_{rec}=1.0\times E_{true}$ & 0.5\% & NuTeV testbeam
measurement\\
& Angular Resolution & $\delta\theta=0.011/E^{0.96}$ rad & 2.5\% & Multiple
scattering fit simulation\\
\multicolumn{2}{l|}{Electron} &  &  & \\
& Energy Resolution & $\delta E/E=0.23/E^{0.5}$ & 1.0\% & Same as CHARM II\\
& Energy Scale Error & $E_{rec}=1.0\times E_{true}$ & 1.0\% & Scaled from
CHARM II with NuSOnG statistics\\
& Angular Resolution & $\delta\theta=0.008/E^{0.5}$ rad & 2.5\% & $2$ better
than CHARM II due to sampling\\
\multicolumn{2}{l|}{Flux} &  &  & \\
& Normalization & 1.0 & 3\% & Current total cross section uncertainty\\
& Shape Uncertainty & 1.0 & 1\% & Similar to NuTeV low-nu method\\
\multicolumn{2}{l|}{Backgrounds} &  &  & \\
& $\nu_\mu$ CCQE & 1.0 & 5\% & Extrapolated from NuTeV\\
& $\nu_e$ CCQE & 1.0 & 3\% & Extrapolated from CHARM II\\\hline
\end{tabular}
\end{ruledtabular}

\caption{Resolutions and systematic uncertainty estimates used in the parameterized
Monte Carlo studies.  The NuTeV estimates are based on Ref. \cite{NuTeVres} 
and the CHARM II
estimates from Ref. \cite{CharmIIdet}.  Units for angles are radians and energies are in GeV.}\label{Resolution_syst_errors}%
\end{table*}%

\subsection{A $0.7\%$ Measurement Goal for the ES to IMD Ratio \label{justify}
}

Achieving 0.7\% precision on the ES/IMD measurement depends on reducing the
backgrounds to an acceptable level without introducing significant systematics
and while maintaining high signal statistics.  Many of the
systematic uncertainties will tend to cancel. The most important background
for both the $\nu$-$e$ neutral current and IMD events comes from charged current
quasi-elastic (CCQE) scatters ($\nu_{e}n\rightarrow pe$ and $\nu_{\mu
}n\rightarrow p\mu$). These background CCQE processes have a much broader
$Q^{2}$ as compared to the signal processes and, therefore, can be partially
eliminated by kinematic cuts on the outgoing muon or electron. Initial cuts on
the scattering angle and energy of the outgoing muon or electron can easily
reduce the CCQE background by factors of 60 and 14 respectively while
retaining over 50\% of the $\nu$-$e$ neutral current and IMD signal. This leaves
events with very forward scatters and outgoing scattered protons of low
kinetic energy.

Because the NuSOnG design is at the conceptual stage and in order to be
conservative, we have developed two different strategies for achieving a 0.7\%
error. This serves as a proof of principle that this level of error, or
better, can be reached. The first method relies on detecting protons from the
quasi-elastic scatter. The second method uses the beam kinematics to cut the
low energy flux which reduces the CCQE background.

These methods were checked via two, independently written, parameterized Monte
Carlos. The parameterized Monte Carlos made the assumptions given in Table
\ref{Resolution_syst_errors} where both the assumed values and uncertainties
are presented. These estimates of resolutions and systematic errors are based
on previous experimental measurements or on fits to simulated data. One Monte
Carlo used the Nuance event generator \cite{Nuance} to produce events, while the other was
an independently written event generator. Both Monte Carlos include nuclear
absorption and binding effects.%

The first strategy uses the number of protons which exit the glass to
constrain the total rate of the background. In $\sim 33\%$ of the
events, a proton will exit the glass, enter a chamber and traverse the
gas. This samples protons of all energies and $Q^{2}$, since the
interactions occur uniformly throughout the glass. After initial cuts,
the protons are below 100 MeV, and therefore highly ionizing.  If
we define 1 MIP as the energy deposited by a single minimum ionizing
particle, like a muon, then the protons 
consistently deposit greater than 5 MIPs in the chamber. Thus, one
can identify CCQE events by requiring
$>$%
4 MIPS in the first chamber. The amount of remaining CCQE background
after this requirement can be measured if a fraction such as 10\% of
the detector is made from scintillating glass that can directly
identify CCQE events from light associated with the outgoing proton. A
wide range of scintillating glasses have been developed \cite{glasses}
for nuclear experiments. These glasses are not commonly used in high
energy physics experiments because the scintillation time constant is
typically on the order of 100 ns. In a neutrino experiment, which 
has inherently lower rates than most particle experiments,
this is not an issue. CCQE events can be identified by the
scintillation light from the proton assuming reasonable parameters for
the glass and readout photomultiplier tubes: 450 photons/MeV, an
attenuation length of 2 m, eight phototubes per glass sheet, quantum
efficiency of the tubes of 20\%. Using the identified CCQE events from
the instrumented glass, the uncertainty in the residual background can
be reduced to 2.0\% for the IMD measurement. For the CCQE background
to the $\nu_\mu $-$e$ neutral current measurement, the uncertainty is
assumed to be 3\% for the Monte Carlo prediction. Combining all the
systematic errors leads to a $\sim $0.7\% accuracy on the $\nu$-$e$
measurement as shown in Tab.~\ref{Method_1_errors}.%

In Tab.~\ref{Method_1_errors}, the cancellation of the flux errors
should be noted.  This occurred because we use the ES/IMD ratio,
as discussed in the previous section.

\begin{table*}[tbp] \centering
\begin{ruledtabular}
\begin{tabular}
[c]{rrccc}\hline
&  & IMD Uncertainty & ES Uncertainty & Uncertainty on Ratio\\\hline
\multicolumn{2}{l}{Statistical Uncertainty} & 0.18\% & 0.46\% & 0.49\%\\\hline
\multicolumn{2}{l}{Resolution Smearing} &  &  & \\
& $\delta$(E$_{\mu}$) $=\pm2.5\%$ & 0.00\% & 0.00\% & 0.00\%\\
& $\delta$($\theta_{\mu}$) $=\pm2.5\%$ & 0.04\% & 0.00\% & 0.04\%\\
& $\delta$(E$_{e}$) $=\pm1.5\%$ & 0.00\% & 0.01\% & 0.01\%\\
& $\delta$($\theta_{e}$) $=\pm2.5\%$ & 0.00\% & 0.09\% & 0.09\%\\\hline
\multicolumn{2}{l}{Energy Scale} &  &  & \\
& $\delta$(Escale$_{\mu}$) $=0.5\%$ & 0.37\% & 0.00\% & 0.37\%\\
& $\delta$(Escale$_{e}$) $=1.5\%$ & 0.00\% & 0.19\% & 0.19\%\\\hline
\multicolumn{2}{l}{Flux} &  &  & \\
& Normalization & 3.00\% & 3.00\% & 0.00\%\\
& High energy flux up 1\% & 0.25\% & 0.25\% & 0.00\%\\
& Low energy flux up 1\% & 0.15\% & 0.13\% & 0.02\%\\\hline
\multicolumn{2}{l}{IMD Background: statistical error} & 0.06\% & 0.00\% &
0.06\%\\
\multicolumn{2}{r}{2.0\% systematic error} & 0.26\% & 0.00\% & 0.26\%\\\hline
\multicolumn{2}{l}{$\nu_{\mu}$e Background: statistical error} & 0.00\% &
0.12\% & 0.12\%\\
\multicolumn{2}{r}{3\% systematic error} & 0.00\% & 0.19\% & 0.19\%\\\hline
&  & \multicolumn{2}{r}{Total Syst. Uncertainty on Ratio} & 0.54\%\\
&  & \multicolumn{2}{r}{Total Stat. Uncertainty on Ratio} & 0.51\%\\
&  & \multicolumn{2}{r}{Total Uncertainty on Ratio} & 0.74\%\\\hline
\end{tabular}
\end{ruledtabular}

\caption{Estimates of the IMD and ES uncertainties using a $>5$ MIP cut on the first downstream
chamber.  The columns give the errors for each process and then for the ratio.  Errors are included for
statistical uncertainties and uncertainties associated with the knowledge of resolution smearing,
energy scale, flux shape, and backgrounds.  The flux shape uncertainties are significantly reduced
in the ratio measurement.}\label{Method_1_errors}%
\end{table*}%

The second strategy involves reducing the relative CCQE background to
signal by using a harder flux for the analysis.  This study used the
same Monte Carlos, with the resolutions listed in
Tab.~\ref{Resolution_syst_errors}, as the first analysis.  The total
systematic and statistical error achieved was 0.6\%.  Below, we
explain how a harder flux is obtained for the analysis.  Then, we
explain how this flux improves the signal-to-background in both the ES
and IMD analyzes.

The strong correlation between energy and angle at the NuSOnG detector
is used to isolate the harder flux. This is simplest to express in 
the non-bend view of the beamline, where it is given for pions by
the well-known off-axis formula:
\begin{equation}
E_\nu = {{0.43 E_{\pi}}\over{1+\gamma^2 \theta^2}},
\end{equation}
where $\theta$ is the off-axis angle, $\gamma=E_\pi/m_\pi$, $E_\pi$ is 
the energy of the pion and $E_\nu$ is the energy of the neutrino.
For the NuTeV beam and detector lay-out, this angle-energy dependence 
resulted in the sharp cutoff of the flux for $<30$ GeV
shown in  Fig. \ref{fig:beam}.   Using the NuTeV G3 beam Monte Carlo \cite{NuTeVbeam},
we have shown that by  selecting vertices in the
central region of the detector, one can adjust the energy where the
flux sharply cuts off.  Adjusting the aperture to retain flux
above 50 GeV reduces the total event rate by 55\%.

A harder flux allows for background reduction in both the ES and the
IMD samples while maintaining the signal at high efficiency.  In the
case of ES events, the background is from $\nu_e$ CCQE.  The energy
distribution of the electron is substantially different in the two
cases.  In the case of $\nu_e$ CCQE events, the electron carries most
of the energy of the incoming neutrino because the exchange energy in
the interaction is small.  Thus the CCQE events produced by the harder
flux populate the visible energy range above 50 GeV.  On the other
hand, the outgoing electron in ES events tends to populate the low
visible energy region due to the combination of a flat $y$
distribution for the process convoluted with the incident neutrino
energy spectrum.  The result is that a cut on the visible energy less
than 50 GeV reduces the error from the $\nu_e$ CCQE background to a
negligible level.  To understand the improvement in the IMD analysis,
consider Fig.~\ref{fig:IMDthresh}, which shows the threshold effects.
The IMD signal is also rising with energy.  In contrast, the $\nu_\mu$
CCQE rate, which is the most significant background, is flat with
energy for fluxes above 1 GeV.  This signal-to-background is greatly
improved with a high energy flux.  This allows looser cuts to be
applied, which in turn reduces the systematics.

These two analyzes use substantially different strategies and can, in
principle, be combined. Given these preliminary studies, we feel
confident that as the detector moves from a conceptual to real design,
we will be able to achieve a better than  0.7\% error.   However,
for this paper we take the conservative approach of assuming 0.7\%.

\subsection{Neutrino Quark Scattering \label{PWsection}}

~~~\\
~~~

Substantially higher precision has been obtained using neutrino-quark
scattering, which compares neutral-current (NC) to charged-current (CC)
scattering to extract $\sin^2 \theta_W$.  However, these experiments are 
subject to issues of modeling in the quark sector.  
Fig.~\ref{pastnu}(bottom) reviews the history of these measurements.

The lowest systematic errors come from 
implementing a ``Paschos-Wolfenstein style'' \cite{PW} analysis.
This PW technique would be 
used by any future experiment, including NuSOnG.  This requires high 
purity 
$\nu$ and $\bar \nu$ beams, for which 
the following ratios of DIS events could be formed:
\begin{eqnarray}
R^\nu &=& \frac{\sigma_{NC}^\nu}{\sigma_{CC}^\nu} \\
R^{\bar \nu} &=& \frac{\sigma_{NC}^{\bar \nu}}{\sigma_{CC}^{\bar \nu}}. 
\end{eqnarray}
Paschos and Wolfenstein \cite{PW} recast these as:
\begin{equation}
R^- = \frac{\sigma_{NC}^\nu - \sigma_{NC}^{\bar \nu}}{\sigma_{CC}^\nu - \sigma_{CC}^{\bar \nu}} = \frac{R^\nu - r R^{\bar \nu}}{1-r},
\label{eq:PWRminus}
\end{equation}
where $r=\sigma_{CC}^{\bar \nu}/\sigma_{CC}^{\nu}$.  In $R^-$ many systematics cancel to first order, including the effects of the quark and
antiquark seas for $u, d, s$, and $c$.  Charm production only enters
through $d_{valence}$ (which is Cabibbo suppressed) and at high $x$;
thus the error from the charm mass is greatly reduced.  
The cross section ratios can be written in terms of the effective 
neutrino-quark coupling parameters $g_L^2$ and $g_R^2$ as
\begin{eqnarray}
R^\nu &=& g_L^2+rg_R^2  \\
R^{\bar \nu} &=& g_L^2 + {1 \over r} g_R^2\\
R^- &=& g_L^2-g_R^2 = \rho^2 ({1 \over 2} - \sin^2\theta_W),
\end{eqnarray}
in which
\begin{eqnarray}
g_L^2 & = &  (2 g_L^\nu g_L^u)^2 + (2 g_L^\nu g_L^d)^2~\cr
& = & \rho^2 ({1 \over 2} - \sin^2 \theta_W + {5 \over 9} \sin^4 \theta_W) \label{eq:gl}\\
g_R^2 & = &  (2 g_L^\nu g_R^u)^2 + (2 g_L^\nu g_R^d)^2 \cr
& = &\rho^2({5 \over 9} \sin^4 \theta_W). \label{eq:gr}
\end{eqnarray}

In a variation on the PW idea, rather than directly form $R^-$, NuTeV
fit simultaneously for $R^\nu$ and $R^{\bar \nu}$ to extract $\sin^2
\theta_W,$ obtaining the value $\sin^2\theta_W=0.2277\pm 0.00162$.
Events were classified according to the length of hits in the
scintillator planes of the NuTeV detector, with long events
identified as CC interactions and short events as NC.  An important
background in the CC sample came from pion decay-in-flight, producing
a muon in a NC shower.  Significant backgrounds in the NC sample came
from muons which ranged out or exited and from $\nu_e$ CC scatters
which do not have a muon and thus are classified as ``short.''

In this paper, we present the sensitivity of NuSOnG to new physics if
the NuTeV errors are reduced by a factor of $\sim 2$.  This is a very
conservative estimate, since most of the improvement comes from higher
statistics.  Only a 90\% improvement in the systematics is required to
reach this goal.  Tab.~\ref{NuTeVerrs} argues why a 90\% reduction in
systematic error should be straightfroward to achieve.  It is likely
that the NuSOnG errors will be lower, but this requires detailed
study.

In Table~\ref{NuTeVerrs}, we list the errors which NuTeV identified in their
original analysis and indicate how NuSOnG will improve each error.
Many of the largest experimental systematics of
NuTeV are improved by introducing a fine-grained sampling calorimeter.
The NuTeV detector had four inches of iron between unsegmented
scintillator planes and eight inches between drift chamber planes.
Better lateral segmentation and transverse detection will 
improve identification of scatters from intrinsic $\nu_e$s in
the beam and separation of CC and NC events by improved 
three-dimensional shower shape
analyzes.  The NuTeV analyzes of the intrinsic $\nu_e$ content \cite{Serge}
and the CC/NC separation for the $\sin^2 \theta_W$ analysis which
relied strictly on event length.      With this said, the power 
of classifying by event length is shown by the fact that the 
NuTeV intrinsic $\nu_e$ analysis was sensitive to a discrepancy in
the predicted intrinsic $\nu_e$ rate which was recently resolved
with a new measurement of the $K_{e3}$ branching ratio that was
published in 2003.     Details of these issues are considered in 
the next section.

\begin{table*}
\begin{ruledtabular}
  \begin{tabular}{|c|c|l|} \hline

 Source & NuTeV & Method of reduction in NuSOnG \\ 
        & Error &     \\ \hline \hline
 Statistics & 0.00135 & Higher statistics \\ \hline \hline $\nu_e$, $\bar \nu_e$ flux prediction & 0.00039 &  Improves in-situ measurement of $\bar \nu_e$ CC scatters, thereby constraining prediction,\\ 
     &  & due to better lateral segmentation and transverse detection.\\ 
     &  & Also, improved beam design to further reduce $\bar \nu_e$ from $K^0$.\\ \hline
    Interaction vertex position & 0.00030 & Better lateral segmentation.  \\ \hline
    Shower length model & 0.00027 & Better lateral segmentation and transverse detection \\
    & & will allow more sophisticated shower identification model. \\ \hline 
Counter efficiency and noise & 0.00023 &  Segmented scintillator strips of the type \\ \hline 
& & developed by MINOS \cite{MINOSStrips} will improve this. \\ \hline
Energy Measurement & 0.00018 & Better lateral segmentation. \\ \hline \hline 
Charm production, strange sea &
    0.00047 & In-situ measurement \cite{EOI, QCDPRD}. \\
\hline 
$R_L$ &
    0.00032 & In-situ measurement \cite{EOI, QCDPRD}.\\ 
\hline 
$\sigma^{\bar
      \nu}/\sigma^{\nu}$ & 0.00022 & Likely to be at a similar level. \\
    \hline Higher Twist & 0.00014 & Recent results reduce this error \cite{Petti}. \\ \hline 
Radiative Corrections & 0.00011 & New analysis underway, see text below. \\ \hline 
Charm Sea & 0.00010 &  Measured in-situ using wrong-sign muon production in DIS.    \\
\hline 
    \hline Non-isoscalar target & 0.00005 & Glass is isoscalar \\
    \hline
\end{tabular}

\caption{Source and value of NuTeV errors on $\sin^2 \theta_W$, and
  reason why the error will be reduced in the PW-style analysis of
  NuSOnG.  This paper assumes NuSOnG will reduce the total NuTeV error
  by a factor of two.  This is achieved largerly through the improved
  statistical precision and requires only a 90\% reduction in the
  overal NuTeV systematic error.  This table argues that a better than
  90\% reduction is likely, but further study, once the detector
  design is complete, is required.}
\label{NuTeVerrs}
\end{ruledtabular}
\end{table*}

\subsection{The NuTeV Anomaly \label{nutevsection}
}

From Fig.~\ref{pastnu}, it is apparent that the NuTeV measurement
is in agreement with past neutrino scattering results, although
these have much larger errors;
however, in disagreement with the
global fits to the electroweak data which give a Standard Model
value of $\sin^2\theta_W =0.2227$ \cite{NuTeVanomaly}.  
Expressed in terms of the couplings,
NuTeV measures:
\begin{eqnarray}
g_L^2 = 0.30005 \pm 0.00137 \\
g_R^2 = 0.03076 \pm 0.00110, 
\end{eqnarray}
which can be compared to the Standard Model values of $g_L^2=0.3042$ and 
$g_R^2=0.0301$, respectively.

NuTeV is one of a set of $Q^2 \ll m_Z^2$ experiments measuring
$\sin^2\theta_W$.  It was performed at $Q^2 =$ 1 to 140 GeV$^2$,
$\langle Q^2_\nu \rangle=26$ GeV$^2$, $\langle Q^2_{\bar \nu} \rangle
=15$ GeV$^2$, which is also the expected range for NuSOnG.  Two other
precision low $Q^2$ measurements are from atomic parity
violation \cite{APV} (APV), which samples $Q^2 \sim 0$; and SLAC E158,
a M{\o}ller scattering experiment at average $Q^2=0.026$ GeV$^2$
\cite{E158}.  Using the measurements at the $Z$-pole with $Q^2=M_z^2$
to fix the value of $\sin^2 \theta_W$, and evolving to low
$Q^2$\cite{E158web}, the APV and SLAC E158 are in agreement with the
Standard Model.  However, the radiative corrections to neutrino
interactions allow sensitivity to high-mass particles which are
complementary to the APV and M{\o}ller-scattering corrections.  Thus,
these results may not be in conflict with NuTeV.  The NuSOnG
measurement will provide valuable additional information on this
question.

Since the NuTeV result was published, more than 300 papers have been
written which cite this result.  
Several
``Standard-Model'' explanations have been suggested.  While 
some constraints on these ideas can come from outside experiments,
it will be necessary for any future neutrino scattering experiment, 
such as NuSOnG,
to be able to directly address these proposed solutions.
Also various Beyond Standard Model
explanations have been put forward; those which best explain the
result require a follow-up experiment which probes the neutral weak
couplings specifically with neutrinos, such as NuSOnG. 
Here, we consider the explanations which are 
``within the Standard Model'' and address the Beyond Standard 
Model later.

\begin{figure}
\vspace{5mm}
\centering
\scalebox{0.4}{\includegraphics[bb=0 155 585 700]{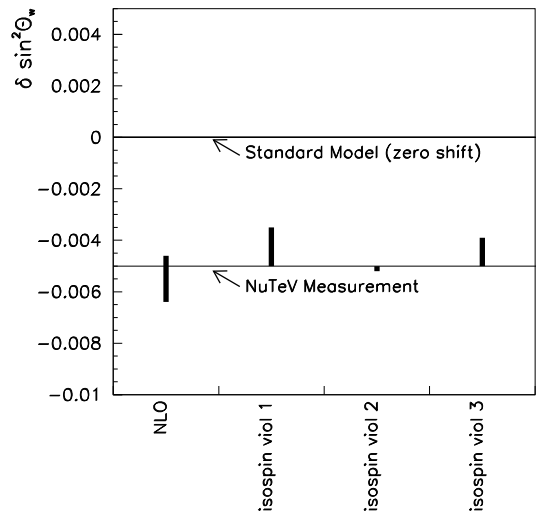}}
\caption{\label{pulls} Effect of various ``Standard Model''
  explanations on the NuTeV anomaly. The $y$-axis is the deviation
  ($\delta
  \sin^2\theta_W=\sin^2\theta_W^{SM}-\sin^2\theta_W^{NuTeV}$).  The
  solid line is the published NuTeV deviation.  Thick black lines
  extending from the NuTeV deviation show the range of possible pulls
  from NLO QCD and various isospin violation models.  Note that the
  isospin violation models are mutually exclusive and so should not be
  added in quadrature.  They are, from left to right, the full bag
  model, the meson cloud model, and the isospin QED model. }
\end{figure}

Several systematic adjustments to the NuTeV result have been
identified since the result was published but have not yet been
incorporated into a new NuTeV analysis.  As discussed here, the
corrections due to the two new inputs, a new $K_{e3}$ branching ratio
and a new strange sea symmetry, are significant in size but are in
opposite direction -- away and toward the Standard Model.  So a
re-analysis can be expected to yield a central value for NuTeV which
will not change significantly.  However, the error is expected to
become larger.

In 2003, a new result from BNL865 \cite{Ke3} yielded a
$K_{e3}$ branching ratio which was $2.3\sigma$ larger than 
past measurements and a value of $|V_{us}|^2$ which brought
the CKM matrix measurements into agreement with unitarity in the
first row \cite{865impact}.    The measurement was confirmed by
CERN NA48/2 \cite{Na48}.
The resulting increased $K_{e3}$ branching ratio \cite{PDG}
increases the absolute prediction of intrinsic $\nu_e$s in the NuTeV
beam.  This does not significantly change the error because the error
on $Ke3$ was already included in the analysis.  However, it introduces
a correction moving the NuTeV result further away from the Standard
Model, since it implies that in the original analysis, NuTeV
under-subtracted the $\nu_e$ background in the NC sample.  The shift in
$\sin^2\theta_W$ can be estimated in a back of envelope calculation to
be about $\sim$0.001 away from the Standard Model
\cite{SamPrivateComm}.

The final  NuTeV
measurement of the difference between the strange and anti-strange sea
momentum distributions, was published in 2007 \cite{DMason}.   This
``strange sea asymmetry'' is defined as 
\begin{equation}
 xs^-(x)\equiv xs(x)-x\overline{s}(x),
\end{equation}
Because of mass suppression for the production of charm in CC scatters
from strange quarks, a difference in the momentum distributions will result
in a difference in the CC cross sections for neutrinos and
antineutrinos. Thus a  correction to the denominator of
Eq.~(\ref{eq:PWRminus}) would be required.  The most recent next-to-leading
order analysis finds the asymmetry, integrated over $x$ is $0.00195\pm
0.00055\pm 0.00138$ \cite{DMason}.  An integrated asymmetry of 0.007
is required to explain the published NuTeV result \cite{DMason}, and so
one can estimate that this is a shift of about 0.0014 in
$\sin^2\theta_W$ toward the Standard Model.  In this case, the errors
on the NuTeV result will become larger because this effect was not
originally considered in the analysis.  A very naive estimate of the
size of the increase can be derived by scaling the error on the
integrated strange sea, quoted above, and is about 0.001 toward
the Standard Model. If this naive estimate of the 
systematic error is borne out, then this could
raise the NuTeV error on $\sin^2\theta_W$ from 0.0016 to 0.0018.
NuSOnG will directly address the strange sea asymmetry in 
its QCD measurement program, as described in ref. \cite{QCDPRD}.

In ref. \cite{diener}, additional electromagnetic radiative
corrections have been suggested as a source of the discrepancy.
However, this paper only considered the effect of these corrections on
$R^\nu$ and not $R^{\bar \nu}$ and for fixed beam energy of $E_\nu=80$
GeV.  The structure of the code from these authors has also made it
difficult to modify for use in NuTeV.  This has prompted a new set of
calculations by other authors which are now under way \cite{newradcor}.
There are, as yet, only estimates for the approximate size of newly
identified effects, which are small.

The NuTeV analysis was not performed at a full NLO level in QCD; any
new experiment, such as NuSOnG will need to undertake a full NLO
analysis.  This is possible given recently published calculations
\cite{Ellis,Moch}, including those on target mass corrections
\cite{SamTargMass}.  On Fig.~\ref{pulls}, we show an early estimate of
the expected size and direction of the pull \cite{nlomodels}.  On this
plot, the solid horizontal line indicates the deviation of NuTeV from
the Standard Model.  The thick vertical lines, which emanate from the
NuTeV deviation, show the range of pulls estimated for various
explanations. The range of pull for the NLO calculation is shown on
the left.

The last possibility is that there is large isospin violation (or
charge symmetry violation) in the nucleus.  The NuTeV analysis assumed
isospin symmetry, that is, $u(x)^p = d(x)^n$ and $d(x)^p = u(x)^n$.
Isospin violation can come about from a variety of sources and is an
interesting physics question in its own right. NuSOnG's direct
constraints on isospin violation are discussed in ref. \cite{QCDPRD}, 
which also considers the constraints from
other experiments.
Various models for isospin violation have
been studied and their pulls range from less than $1\sigma$ away from
the Standard Model to $\sim 1 \sigma$ toward the Standard Model
\cite{isomodels}. We have chosen three examples \cite{isomodels} for
illustration on Fig.~\ref{pulls}: the full bag model, the meson cloud
model, and the isospin QED model.  These are mutually exclusive
models, so only one of these can affect the NuTeV anomaly.


\section{The Terascale Physics Reach of NuSOnG\label{Terascale}}

\begin{center}
\begin{table*}[tbp] 
\begin{ruledtabular}
\begin{tabular}{|l|l|}
  Topic &   Contribution of NuSOnG Measurement  \\ \hline \hline
  Oblique Corrections &   Four distinct and complementary probes of $S$ and $T$.\\ 
  &    In the case of agreement with LEP/SLD: $\sim$25\% improvement in 
  electroweak precision. \\ \hline
  Neutrino-lepton NSIs  &  Order of magnitude improvement in neutrino-electron effective couplings measurements.\\
  &  Energy scale sensitivity up to $\sim 5$ TeV at 95\% CL.\\ \hline
Neutrino-quark NSIs   &  Factor of two improvement in neutrino-quark effective coupling measurements.\\ 
  &  Energy scale sensitivity up to $\sim 7$ TeV at 95\% CL.\\ \hline
  Mixing with Neutrissimos &  30\% improvement on the $e$-family coupling in a global fit.  \\ 
 & 75\% improvement on the $\mu$-family coupling in a global fit.  \\ \hline
  Right-handed Couplings &  Complementary sensitivity to $g_R/g_L$ compared to LEP.\\
 & Order of magnitude improvement compared to past experiments. \\
\end{tabular}
\end{ruledtabular}
\caption{Summary of NuSOnG's contribution to general Terascale physics studies.}
\label{Tab:genericsummary}
\end{table*}
\end{center}

Even when new states are too heavy to be produced at resonance in
collisions they can make their presence known indirectly, as virtual
particles which affect SM processes through interference with SM
contributions to amplitudes.  The new heavy states induce small shifts
in observables from SM predictions, and conversely precise
measurements of these observables can constrain or detect new physics
at mass scales well above the energies of the colliding particles.  In
this way the precision neutrino scattering measurements at NuSOnG will
place TeV-scale indirect constraints on many classes of new physics,
or perhaps detect new physics by measuring deviations from SM
predictions.  The effects of new high-scale physics may be reduced to
a small number of effective operators along with corresponding
parameters which may be fit to data.  Although the particular set of
operators used depends on broad assumptions about the new physics, the
approach gives a parameterization of new physics which is largely
model-independent.

For concreteness we will assume that NuSOnG will be able to measure
the neutrino ES/IMD ratio to a precision of 0.7\%, $\sigma({\bar
  \nu}_\mu e)$ (normalized as per Sec.~\ref{IMDsub}) to 1.3\%, and
that NuSOnG will be able to halve the errors on NuTeV's measurement of
DIS effective couplings, to $\Delta g_L^2=0.0007$ and $\Delta
g_R^2=0.0006$ (where $g_L$ and $g_R$ were defined in Eqs.~(\ref{eq:gl})
and (\ref{eq:gr})).

We first parameterize new physics using the oblique parameters $ST$,
which is appropriate when the important effects of the new physics
appear in vacuum polarizations of gauge bosons.  We next assume new
physics effects manifest as higher-dimensional operators made of SM
fermion fields.  We separately consider the possibility that the gauge
couplings to neutrinos are modified.  Realistic models usually
introduce several new operators with relations among the coefficients;
we consider several examples.  A summary of the contributions of
NuSOnG to the study of Terascale Physics is provided in
Table~\ref{Tab:genericsummary}.

\subsection{Oblique corrections \label{obliquecorrections}}

For models of new physics in which the dominant loop corrections are
vacuum polarization corrections to the $SU(2)_L \times U(1)_Y$ gauge
boson propagators (``oblique'' corrections), the $STU$
\cite{Peskin:1991sw, Peskin:1990zt} parameterization provides a
convenient framework in which to describe the effects of new physics
on precision electroweak data.  Differences between the predictions of
a new physics model and those of a reference Standard Model (with a
specified Higgs boson and top quark mass) can be expressed as nonzero
values of the oblique correction parameters $S$, $T$ and $U$.  $T$ and
$U$ are sensitive to new physics that violates isospin, while $S$ is
sensitive to isospin-conserving physics.  Predictions of a Standard
Model with Higgs or top masses different from the reference Standard
Model may also be subsumed into shifts in $S$ and $T$ (in many models
$U$ is much smaller than $S$ and $T$ and is largely unaffected by the
Higgs mass, so it is often omitted in fits).  Within a specific model
of new physics the shift on the $ST$ plot away from the SM will be
calculable \cite{Peskin:2001rw}.  For example,
\begin{itemize}
\item A heavy Standard Model Higgs boson will make a positive
contribution to $S$ and a larger negative contribution to $T$.
\item Within the space of $Z^\prime$ models, a shift in almost any direction
in $ST$ space is possible, with larger shifts for smaller $Z^\prime$
masses.   
\item Models with a fourth-generation of
fermions will shift $S$ positive, and will shift $T$ positive if there
are violations of isospin.  
\end{itemize}
In constructing models incorporating
several types of new physics the corresponding shifts to $S$ and $T$
combine; if contributions from different sectors are large, then they must
conspire to cancel.

\begin{figure}[ht]
\includegraphics[scale=0.4]{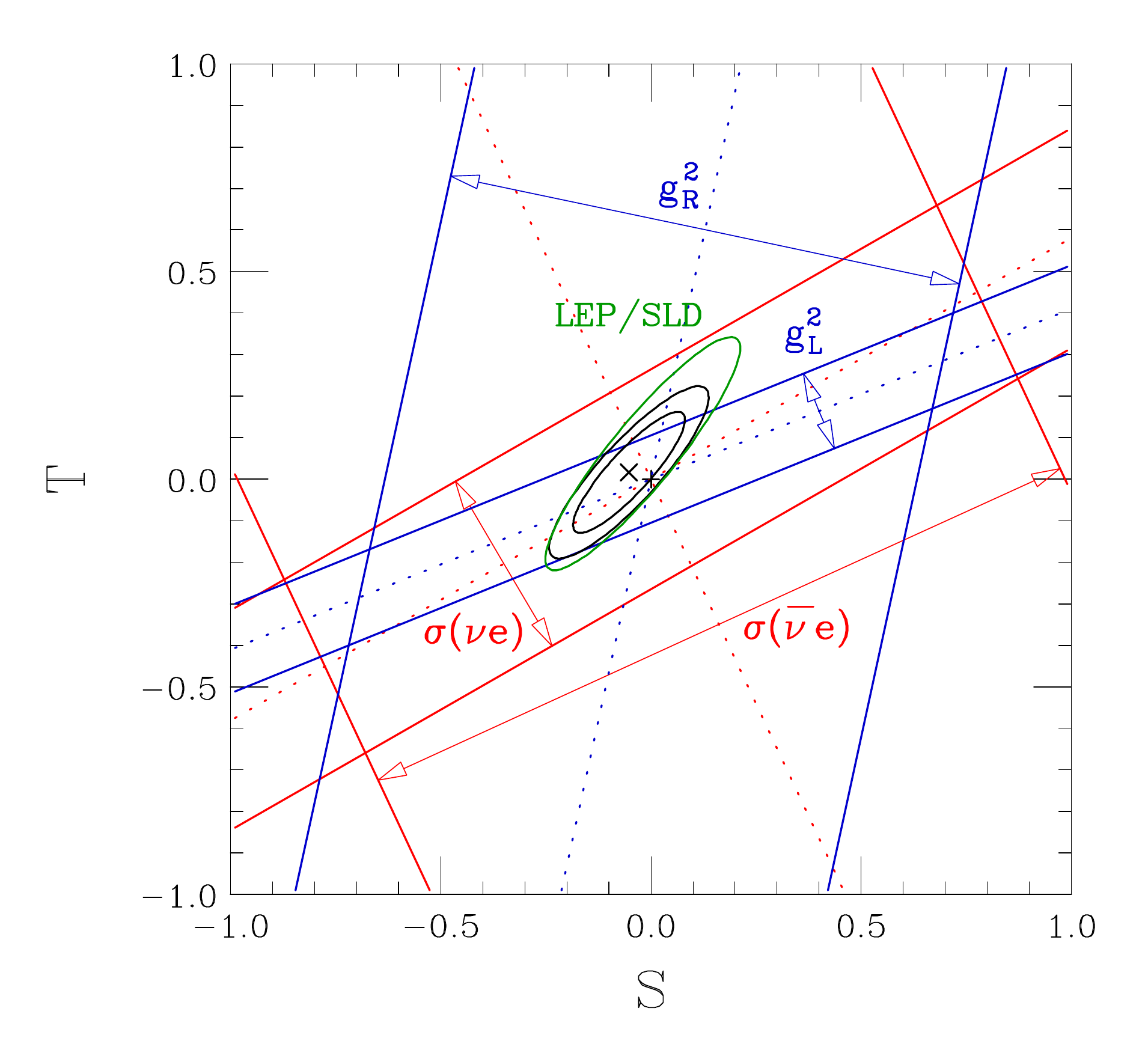}
\caption{The impact of NuSOnG on the limits of $S$ and $T$.
The reference SM is $m_t=170.9$~GeV, and $m_H=115$~GeV.
1$\sigma$ bands due to NuSOnG observables are shown against the 90\%
contour from LEP/SLD.  The central ellipses are the 68\% and 90\% confidence limit contours with NuSOnG included.  See Eqs.~(\ref{eq:gl}) and (\ref{eq:gr}) for the definitions
of $g_L$ and $g_R$.}
\label{STplot1}
\end{figure}

The constraints on $S$ and $T$ from 
the full set of precision electroweak data 
strongly restrict the models of new physics which are viable. 
The strongest constraints are from LEP/SLD,
which give a current bound of 
\begin{eqnarray}
S & = & -0.02\pm 0.11 \;,\cr
T & = & +0.06\pm 0.13 \;,\cr
\mathrm{Corr}(S,T) & = & 0.91.\;
\end{eqnarray}
The ES and DIS measurements from NuSOnG provide four distinct and 
complementary probes of $S$ and $T$,
as shown in Fig.~\ref{STplot1}.  
If the target precision is achieved, and assuming the NuSOnG agree with SM 
predictions, NuSOnG will further 
reduce the errors on $S$ and $T$ from the LEP/SLD values to
\begin{eqnarray}
S & = & -0.05\pm 0.09 \;,\cr
T & = & +0.02\pm 0.10 \;,\cr
\mathrm{Corr}(S,T) & = & 0.87\;.
\end{eqnarray}
The $\sim 25\%$ reduction in the errors is primarily due to the
improved measurement of $g_L^2$.  We note that the error $g_L^2$ is
likely to be further reduced (see Sec.~\ref{PWsection}), and so the this is
conservative estimate of NuSOnG's contribution to the physics.

\subsection{Non-standard interactions}

NuSOnG will probe new physics that modifies neutrino-quark and
neutrino-electron scattering.  If the masses associated to the new
degrees of freedom are much larger than the center of mass energy
($s=2m_eE_{\rm beam}\lesssim 0.5$~GeV$^2$) then modifications to these
processes are well-described by higher-dimensional effective
operators.  In the context of neutrino reactions, these operators are
also referred to as non-standard interactions (NSI's).  In a
model-independent effective Lagrangian approach these effective
operators are added to the SM effective Lagrangian with arbitrary
coefficients.  Expressions for experimental observables can be
computed using the new effective Lagrangian, and the arbitrary
coefficients can then be constrained by fitting to data.  Typically,
bounds on the magnitude of the coefficients are obtained using only one or a
few of the available effective operators.  This approach simplifies
the analysis and gives an indication of the scale of constraints,
although we must be mindful of relationships among different operators 
that will be imposed
by specific assumptions regarding the underlying physics.

To assess the sensitivity of NuSOnG to ``heavy'' new physics in
neutral current processes, we introduce the following effective
Lagrangian for neutrino-fermion interactions
\cite{Bandyopadhyay:2007kx,Davidson:2003ha,Mohapatra:2005wg}:
\begin{eqnarray}
\mathcal{L}_\mathrm{NSI}
& = &
-\sqrt{2}G_F
\Bigl[\, \bar{\nu}_\alpha \gamma_\sigma P_L \nu_\beta 
\,\Bigr]
\Bigl[\,
 \varepsilon_{\alpha\beta}^{fV}\, \bar{f}\gamma^\sigma f
-\varepsilon_{\alpha\beta}^{fA}\, \bar{f}\gamma^\sigma\gamma_5 f
\,\Bigr] \cr
& = &
-2\sqrt{2}G_F
\Bigl[\, \bar{\nu}_\alpha \gamma_\sigma P_L \nu_\beta 
\,\Bigr]
\Bigl[\,
 \varepsilon_{\alpha\beta}^{fL}\, \bar{f}\gamma^\sigma P_L f\cr
&& ~~~~~~~~~~~~~~~~~~~~~~~~~~~~~~+\varepsilon_{\alpha\beta}^{fR}\, \bar{f}\gamma^\sigma P_R f
\,\Bigr] \;.
\label{NSI}
\end{eqnarray}
where $\alpha,\beta=e,\mu,\tau$ and $L,R$ represent left-chiral and
right-chiral fermion fields.  If $\alpha\neq\beta$, then the
$\alpha\leftrightarrow \beta$ terms must be Hermitian conjugates of
each other, {\it i.e.}
$\varepsilon_{\beta\alpha}=\varepsilon_{\alpha\beta}^*$.  NuSOnG is
sensitive to the $\beta=\mu$ couplings.  This effective Lagrangian is
appropriate for parameterizing corrections to neutral current
processes; an analysis of corrections to charged-current processes
requires a different set of four-fermion operators.

Assuming $\varepsilon_{\alpha\beta}=0$ for $\alpha\neq\beta$
we need consider only the terms $\varepsilon_{\mu\mu}^{f*}$
($*=V,A,L,R$).   If we rewrite Eq.~(\ref{eq:geneffint}) as
\begin{eqnarray}
& & \cr
\mathcal{L} & = &
-\sqrt{2}G_F
\Bigl[\, \bar{\nu}\gamma_\mu P_L \nu 
\,\Bigr]
\Bigl[\,
 g_V^{\nu f}\, \bar{f}\gamma^\mu f
-g_A^{\nu f}\, \bar{f}\gamma^\mu\gamma_5 f
\,\Bigr] \cr
& = &
-2\sqrt{2}G_F
\Bigl[\, \bar{\nu}\gamma_\mu P_L \nu 
\,\Bigr]
\Bigl[\,
 g_L^{\nu f}\, \bar{f}\gamma^\mu P_L f \cr
&& ~~~~~~~~~~~~~~~~~~~~~~~~~~~~+g_R^{\nu f}\, \bar{f}\gamma^\mu P_R f
\,\Bigr] \;,
\label{L2}
\end{eqnarray} 
where
\begin{eqnarray}
g_V^{\nu f} & = & 2g_L^\nu g_V^f 
\;=\; \rho\left( I_3^f - 2 Q^f\sin^2\theta_W \right) \;,\cr
g_A^{\nu f} & = & 2g_L^\nu g_A^f 
\;=\; \rho\left( I_3^f \right) \;,\cr
g_L^{\nu f} & = & 2g_L^\nu g_L^f 
\;=\; \rho\left( I_3^f - Q^f\sin^2\theta_W \right)\;,\cr
g_R^{\nu f} & = & 2g_L^\nu g_R^f 
\;=\; \rho\left( -Q^f\sin^2\theta_W \right)\;,
\end{eqnarray}
then we see that adding Eq.~(\ref{NSI}) to the SM Lagrangian will simply 
shift the effective couplings:
\begin{eqnarray}
g_{V}^{\nu f} & \longrightarrow & 
\tilde{g}_{V}^{\nu f} \;=\; g_{V}^{\nu f} + \varepsilon_{\mu\mu}^{fV}\;,\cr
g_{A}^{\nu f} & \longrightarrow & 
\tilde{g}_{A}^{\nu f} \;=\; g_{A}^{\nu f} + \varepsilon_{\mu\mu}^{fA}\;,\cr
g_{L}^{\nu f} & \longrightarrow & 
\tilde{g}_{L}^{\nu f} \;=\; g_{L}^{\nu f} + \varepsilon_{\mu\mu}^{fL}\;,\cr
g_{R}^{\nu f} & \longrightarrow & 
\tilde{g}_{R}^{\nu f} \;=\; g_{R}^{\nu f} + \varepsilon_{\mu\mu}^{fR}.\;
\end{eqnarray}
Consequently, errors on the $g_{P}^{\nu f}$'s translate directly into
errors on the $\varepsilon_{\mu\mu}^{fP}$'s, $P=V,A$ or $P=L,R$.

\subsubsection{Neutrino-lepton NSI} 

A useful review of present constraints on non-standard
neutrino-electron interactions can be found in ref.
\cite{Barranco}.  As this paper states, and as we show below, an
improved measurement of neutrino-elecron scattering is needed.

The world average value for neutrino-electron effective couplings, dominated by CHARM II, is
\begin{eqnarray}
g_V^{\nu e} & = & -0.040\pm 0.015 \;,\cr
g_A^{\nu e} & = & -0.507\pm 0.014 \;,\cr
\mathrm{Corr}(g_V^{\nu e},g_A^{\nu e}) & = & -0.05\;.
\label{CHARM2result}
\end{eqnarray}
The current $1\sigma$ bounds from CHARM II, Eq.~(\ref{CHARM2result}) 
translates to $|\varepsilon_{\mu\mu}^{eP}|<0.01$, $(P=L,R)$ 
with a correlation of $0.07$~\cite{Bandyopadhyay:2007kx}. 
At the current precision goals, NuSOnG's $\nu_\mu e$ and ${\overline \nu}_\mu e$ will significantly reduce the uncertainties on these NSI's, to
\begin{eqnarray}
|\varepsilon_{\mu\mu}^{eV}| & < & 0.0036\;,\cr
|\varepsilon_{\mu\mu}^{eA}| & < & 0.0019\;,\cr
\mathrm{Corr}(\varepsilon_{\mu\mu}^{eV},\varepsilon_{\mu\mu}^{eA}) & = & -0.57\;,
\label{EPSeVAbounds}
\end{eqnarray}
or in terms of the chiral couplings,
\begin{eqnarray}
|\varepsilon_{\mu\mu}^{eL}| & < & 0.0015\;,\cr
|\varepsilon_{\mu\mu}^{eR}| & < & 0.0025\;,\cr
\mathrm{Corr}(\varepsilon_{\mu\mu}^{eL},\varepsilon_{\mu\mu}^{eR}) & = & 0.64.\;.
\label{EPSePbounds}
\end{eqnarray}

Even in the absence of a  $\sigma(\bar{\nu}_\mu e)$ measurement
$\varepsilon_{\mu\mu}^{eL}$ and $\varepsilon_{\mu\mu}^{eR}$ can be 
constrained from the $\nu_\mu e$ scattering data alone through a fit to 
the recoil electron energy spectrum (see Eq.~(\ref{differentialcrosssection})).

We first consider the constraint on $\varepsilon_{\mu\mu}^{eL}$ and
$\varepsilon_{\mu\mu}^{eR}$ from the total cross section
$\sigma(\nu_\mu e)$.  It is convenient to recast the effective
interaction slightly, as
\begin{widetext}
\begin{eqnarray}
\mathcal{L}_\mathrm{NSI}^{e} & = &
-2\sqrt{2}G_F
\Bigl[\, \bar{\nu}_\alpha \gamma_\sigma P_L \nu_\mu 
\,\Bigr]
\Bigl[\,
 \varepsilon_{\alpha\mu}^{eL}\, \bar{e}\gamma^\sigma P_L e
+\varepsilon_{\alpha\mu}^{eR}\, \bar{e}\gamma^\sigma P_R e
\,\Bigr]  \cr
& = & +\frac{\sqrt{2}}{\Lambda^2}
\Bigl[\, \bar{\nu}_\alpha \gamma_\sigma P_L \nu_\mu 
\,\Bigr]
\Bigl[\,
 \cos\theta\, \bar{e}\gamma^\sigma P_L e
+\sin\theta\, \bar{e}\gamma^\sigma P_R e
\,\Bigr]  \;.
\label{eq:leff}
\end{eqnarray}
\end{widetext}
The new physics is parameterized by two coefficients $\Lambda$ and
$\theta$. $\Lambda$ represents the broadly-defined new physics scale
while $\theta\in[0,2\pi]$ defines the relative coupling of left-chiral
and right-chiral electrons to the new physics. As an example, a
scenario with a purely ``left-handed'' $Z^{\prime}$ that couples to
leptons with coupling $g'$ would be described by $\Lambda \propto
M_{Z'}/g'$ and $\theta=0$ or $\theta=\pi$, depending on the relative
sign between $g'$ and the electroweak couplings.  $\Lambda$ and
$\theta$ are related to to the NSI parameters in Eq.~(\ref{NSI}) by
\begin{equation}
\varepsilon_{\alpha\mu}^{eL} = -\frac{\cos\theta}{2 G_F\Lambda^2}\;,~~~
\varepsilon_{\alpha\mu}^{eR} = -\frac{\sin\theta}{2 G_F\Lambda^2}.\;
\end{equation}
Note that we have generalized from our assumption of the previous
section and not taken $\alpha=\mu$ necessarily.  At NuSOnG, new physics modifies
(pseudo)elastic neutrino--electron scattering.  Here we use the word
``pseudo'' to refer to the fact that we cannot identify the flavor of
the final-state neutrino, which could be different from the incoming
neutrino flavor in the case of flavor changing neutral currents.

The shift in the total cross section is
\begin{eqnarray}
\frac{\delta\sigma(\nu_{\mu}e)}{\sigma(\nu_\mu e)}
& = &
\frac{ \left\{\, 2\, g_L^{\nu e}\, \varepsilon_{\mu\mu}^{eL}
+ (\varepsilon_{\mu\mu}^{eL})^2 \right\}
+ \frac{1}{3}
\left\{\, 2\,g_R^{\nu e}\, \varepsilon_{\mu\mu}^{eR}
+ (\varepsilon_{\mu\mu}^{eR})^2 \right\} }
{ (g_L^{\nu e})^2 + \frac{1}{3}(g_R^{\nu e})^2 }  \cr
& \approx & -\left(\frac{516\,\mathrm{GeV}}{\Lambda}\right)^2
\cos(\theta-\phi) \cr
&& + 0.096\left(\frac{516\,\mathrm{GeV}}{\Lambda}\right)^4
(1+2\cos^2\theta) \;.
\label{nueshift2}
\end{eqnarray}
where
\begin{equation}
\tan\phi\;=\;\frac{g_R^{\nu e}}{3g_L^{\nu e}}\;\approx\;-0.28\;.
\end{equation}
When ${\cal O}(\varepsilon^2)$ terms are negligible, a $0.7\%$ measurement
of $\sigma(\nu_\mu e)$ translates into a
95\% confidence level bound of
\begin{equation}
\Lambda \;>\; (4.4\,\mathrm{TeV})\times\sqrt{|\cos(\theta-\phi)|}\;
\end{equation}
from elastic scattering.

\begin{figure}
\centering
\scalebox{0.35}{\includegraphics[clip=true]{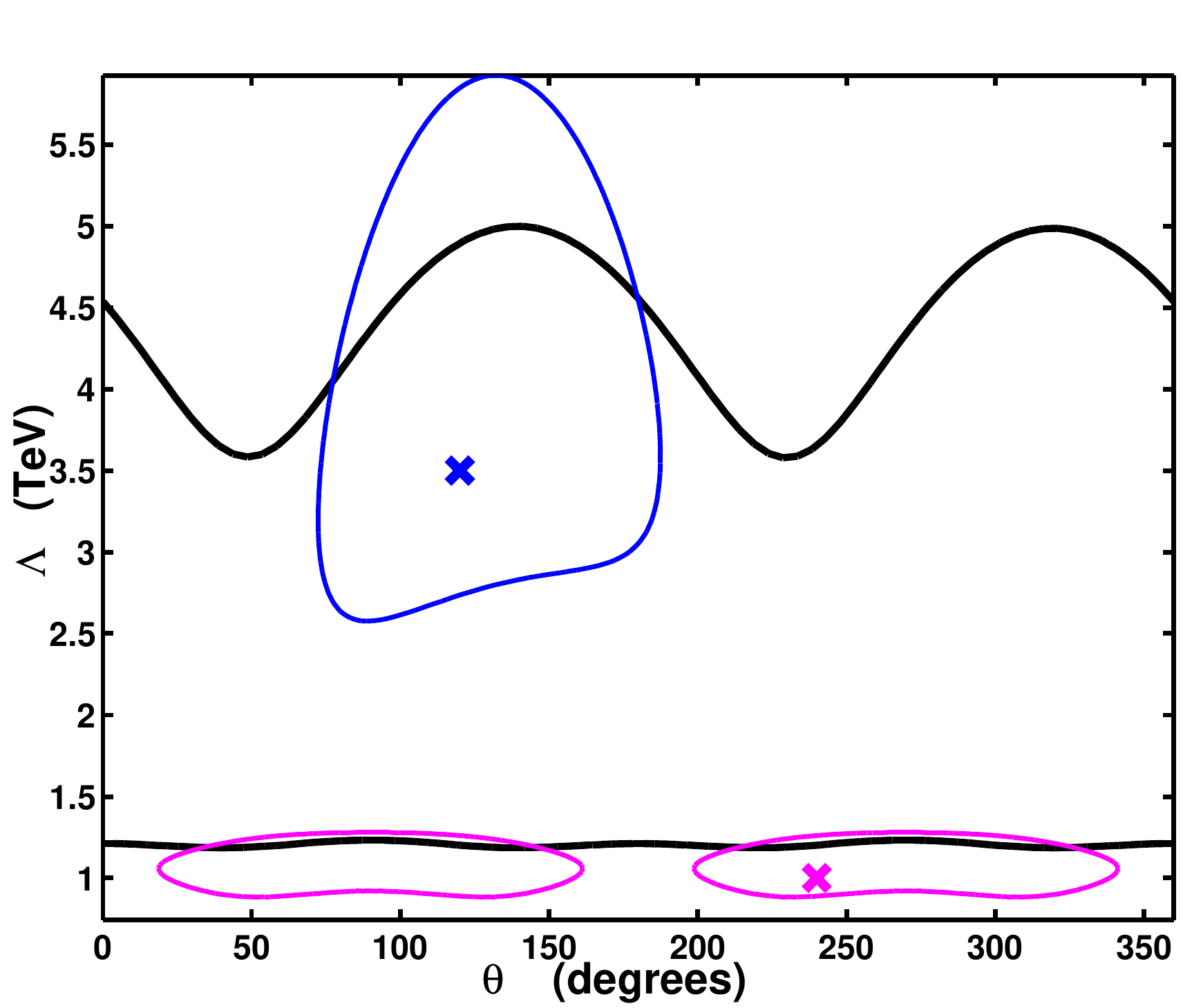}}
\vspace{1mm}
\caption{
(DARK LINES) 95\% confidence level sensitivity of NuSOnG to new heavy  
physics described by Eq.~(\ref{eq:leff}) when $\nu_\alpha=\nu_{\mu}$  
(higher curve) and $\nu_\alpha\neq\nu_{\mu}$ (lower curve).  (CLOSED  
CONTOURS) NuSOnG measurement of $\Lambda$ and  $\theta$, at the 95\%  
level, assuming $\nu_\alpha=\nu_{\mu}$, $\Lambda=3.5$~TeV and $ 
\theta=2\pi/3$ (higher, solid contour) and $\nu_\alpha\neq\nu_{\mu}$, $ 
\Lambda=1$~TeV and $\theta=4\pi/3$ (lower, dashed contour). Note that  
in the pseudoelastic scattering case ($\nu_\alpha\neq\nu_{\mu}$) $ 
\theta$ and $\pi+\theta$ are physically indistinguishable.
}
\label{fig:eff_limit}
\end{figure}

The measurement of the electron recoil energy will allow us to do better.
Fig.~\ref{fig:eff_limit}(dark line) depicts the 95\% confidence level
sensitivity of NuSOnG to the physics described by Eq.~(\ref{eq:leff})
when $\nu_\alpha=\nu_{\mu}$, obtained after fitting the recoil
electron kinetic energy distribution. 
Fig.~\ref{fig:eff_limit}(closed contour) represents how well NuSOnG
should be able to measure $\Lambda$ and $\theta$, at the 95\%
level. Weaker bounds from pseudoelastic scattering are also shown.
We have not included ``data'' from
$\bar{\nu}_{\mu}$--electron scattering. While there will be fewer of these events, 
they should qualitatively improve our
ability to pin down the new physics parameters given the distinct
dependency on $g^{\nu e}_V$ and $g^{\nu e}_A$ (see Sec.~\ref{nuesub}).

Eq.~(\ref{eq:leff}) does not include all effective dimension-six
operators that contribute to neutrino--electron (pseudo) elastic
scattering. All neglected terms will either not contribute at NuSOnG,
or were assumed to be suppressed with respect to
Eq.~(\ref{eq:leff}). In turn, terms proportional to a right-handed
neutrino current $\bar{\nu}_R\gamma_{\sigma}\nu_R$ lead to negligibly
small effects since neutrino masses are negligibly small and we are
dealing with neutrino beams produced by pion and muon decay ({\it
  i.e.}, for all practical purposes, we have a purely left-handed muon
neutrino beam and a purely right-handed muon antineutrino
beam). Chirality violating effective operators
(e.g. $(\bar{\nu}_R\nu_L)(\bar{e}_Le_R)$), on the other hand, are
expected to be suppressed with respect to Eq.~(\ref{eq:leff}) by terms
proportional to neutrino masses and the electron mass (measured in
units of $\Lambda$). The reason is that, in the limit of massless
neutrinos or a massless electron, chiral symmetry is restored while
such operators explicitly violate it. For the same reason,
dimension-five magnetic moment-type operators
($\bar{\nu}\sigma_{\rho\sigma}\nu F^{\rho\sigma}$) have also been
neglected.

We note also that Eq.~(\ref{eq:leff}) violates
$SU(2)_L$ unless one also includes similar terms where $\nu_
L\leftrightarrow\ell_L$ ($\ell=e,\mu, \tau$). 
In this case, certain
flavor combinations would be severely constrained by
electron--electron scattering and rare muon and tau decays. One way
around such constraints is to postulate that the operators in
Eq.~(\ref{eq:leff}) are dimension-eight operators proportional to
$\bar{L}H^*\gamma_{\sigma}LH$, where $L$ is the left-chiral lepton
doublet and $H$ is the Higgs scalar doublet. In this case,
$1/\Lambda^2$ should be replaced by $v^2/\Lambda^4$, where $v=246$~GeV
is the scale of electroweak symmetry breaking.

Finally, another concern is whether modifications to the charged current neutrino--electron 
(pseudo)quasi-elastic scattering  ((pseudo)IMD, $\nu_{\mu}e\to\nu_{\alpha}\mu$) can render the translation of NuSOnG data into constraints or measurements of $\theta$ and $\Lambda$ less straightforward. This turns out not to be the case, since new physics contributions to $\nu_{\mu}e\to\nu_{\alpha}\mu$ are already very well constrained by precision studies of muon decay. Hence, given the provisos of the two previous paragraph, Eq.~(\ref{eq:leff}) is expected to capture all ``heavy'' new physics effects in (pseudo)elastic neutrino electron scattering.

\subsubsection{Neutrino-quark NSI} 

We next consider the $f=u,d$ case. The change in the parameters
$g_L^2$ and $g_R^2$ (see Eqs.~(\ref{eq:gl},\ref{eq:gr})) due to the NSI's is
\begin{eqnarray}
\Delta g_L^2 & = &  2 g_L^{\nu u}\varepsilon_{\mu\mu}^{uL}
+2 g_L^{\nu d}\varepsilon_{\mu\mu}^{dL} \cr
&\;\approx\;& +0.69\,\varepsilon_{\mu\mu}^{uL} -0.85\,\varepsilon_{\mu\mu}^{dL}
\;,\cr
\Delta g_R^2 & = & 
 2 g_R^{\nu u}\varepsilon_{\mu\mu}^{uR}
+2 g_R^{\nu d}\varepsilon_{\mu\mu}^{dR} \cr
&\;\approx\;& -0.31\,\varepsilon_{\mu\mu}^{uR} +0.15\,\varepsilon_{\mu\mu}^{dR}
\;.
\end{eqnarray}
so only these linear combinations are constrained.
The bounds from NuTeV (rescaled to $1 \sigma$ bounds from ref.~\cite{Bandyopadhyay:2007kx}) are:
\begin{eqnarray}
\varepsilon_{\mu\mu}^{uL} & = & -0.0053\pm 0.0020\;,\cr
\varepsilon_{\mu\mu}^{dL} & = & +0.0043\pm 0.0016\;,\cr
|\varepsilon_{\mu\mu}^{uR}| & < & 0.0035\;,\cr
|\varepsilon_{\mu\mu}^{dR}| & < & 0.0073\;.
\end{eqnarray}
These bounds are obtained by setting only one of the parameters be non-zero at a time.
If NuSOnG reduces the errors on the NuTeV measurement of $g_L^2$ and $g_R^2$ by a factor of 2, the $1\sigma$ bounds on the NSI parameters are similarly reduced:
\begin{eqnarray}
|\varepsilon_{\mu\mu}^{uL}| & < & 0.001\;,\cr
|\varepsilon_{\mu\mu}^{dL}| & < & 0.0008\;, \cr
|\varepsilon_{\mu\mu}^{uR}| & < & 0.002\;, \cr
|\varepsilon_{\mu\mu}^{dR}| & < & 0.004\;. 
\end{eqnarray}
In terms of a new physics scale defined as $\Lambda = 1/\sqrt{2 \,\mathrm{G_F} \varepsilon},$ these constraints range from 
$\Lambda \;>\; 3\,\mathrm{TeV}$ to $\Lambda \;>\; 7\,\mathrm{TeV}.$

We note that neutrino-quark scattering will also be sensitive to NSIs which
correct CC interactions.  These interactions are not included in Eq.~(\ref{NSI}). If they
are important, as is the case in some of the scenarios we treat later, a new
analysis is necessary and the bounds above cannot be used. This is to be contrasted to the neutrino--lepton case, discussed in the previous subsection.

\subsection{Neutrissimos, Neutrino Mixing and Gauge Couplings}

\begin{figure}[ht]
\vspace{-1.5in}
\includegraphics[scale=0.4]{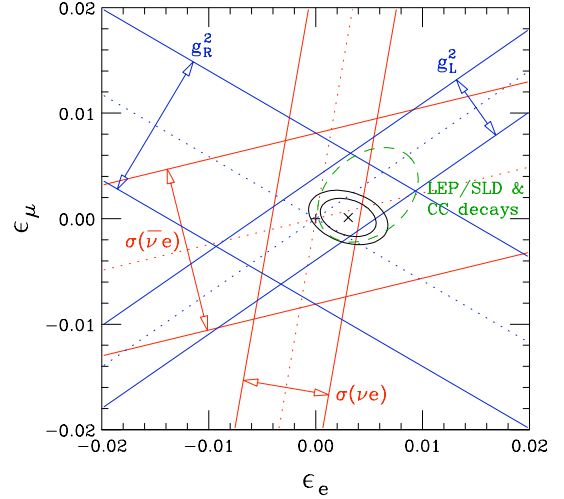}
\caption{Potential constraint on $\epsilon_e$ and $\epsilon_\mu$ from NuSOnG (see Eq.~(\ref{epsilon_ell})).
This is a two-dimensional projection of a 4 parameter fit with $S$, $T$, $\epsilon_e$ and $\epsilon_\mu$.
The green ellipse is the 90\% CL contour of a fit to all the charge current particle decay data + LEP/SLD.}
\label{epsilon_fit}
\end{figure}

In those classes of models which include moderately heavy electroweak
gauge singlet (``neutrissimo'') states, with masses above 45 GeV, the
mixing of the $SU(2)_L$-active neutrinos and the sterile states may
lead to a suppression of the neutrino-gauge couplings.  The resulting
pattern of modified interactions is distinct from those of the
previous section since they will also induce correlated shifts to the
charged-current coupling.  For example, Ref.~\cite{Loinaz:2003gc}
presents models with one sterile state per active neutrino flavor and
intergenerational mixing among neutrinos.  In these models the flavor
eigenstates are linear combinations of mass eigenstates, and those
mass eigenstates too heavy to be produced in final states result in
an effective suppression of the neutrino-gauge boson coupling.  This
suppression may be flavor-dependent depending on the structure of the
neutrino mixing matrix.  If the mass matrix contains Majorana terms,
such models permit both lepton flavor violation and lepton
universality violation.

Neutrinos couple to the $W$ and the $Z$ through interactions described by:
\begin{eqnarray}
\mathcal{L} &\;=\; &
\frac{g}{\sqrt{2}}W^-_\mu\, \bar{\ell}_L \gamma^\mu \nu_{\ell L}
+ \frac{g}{\sqrt{2}}W^+_\mu\, \bar{\nu}_{\ell L} \gamma^\mu \ell_L \cr
& & + \frac{e}{2sc}Z_\mu\, \bar{\nu}_{\ell L}\gamma^\mu \nu_{\ell L} \;,
\end{eqnarray}
where $\ell = e,\mu,\tau$.
If the neutrinos mix with gauge singlet states so that the $SU(2)_L$ interaction eigenstate is a superposition of mass eigenstates $\nu_\mathrm{\ell,light}$ and $\nu_\mathrm{\ell,heavy}$
\begin{equation}
\nu_{\ell L} \;=\; 
\nu_\mathrm{\ell,light}\cos\theta_\ell + \nu_\mathrm{\ell,heavy}\sin\theta_\ell\;,
\end{equation}
then the interaction of the light states is given by
\begin{eqnarray}
\mathcal{L} &&\;=\; \cr
&&\left(\frac{g}{\sqrt{2}}W^-_\mu\, \bar{\ell}_L \gamma^\mu \nu_\mathrm{\ell,light}
+ \frac{g}{\sqrt{2}}W^+_\mu\, \bar{\nu}_\mathrm{\ell,light} \gamma^\mu \ell_L
\right)\cos\theta_\ell \cr
&&+ \left(\frac{e}{2sc}Z_\mu\, \bar{\nu}_\mathrm{\ell,light}\gamma^\mu \nu_\mathrm{\ell,light}\right)\cos^2\theta_\ell\;.
\end{eqnarray}
Defining
\begin{equation}
\epsilon_\ell \;\equiv\; 1-\cos^2\theta_\ell\;.
\end{equation}
the shift in the Lagrangian due to this mixing is
\begin{eqnarray}
\delta\mathcal{L} &\;=\;&
-\left(\frac{g}{\sqrt{2}}W^-_\mu\, \bar{\ell}_L \gamma^\mu \nu_{\ell}
+ \frac{g}{\sqrt{2}}W^+_\mu\, \bar{\nu}_{\ell} \gamma^\mu \ell_L
\right)\frac{\epsilon_\ell}{2} \nonumber \\ 
&-& \left(\frac{e}{2sc}Z_\mu\, \bar{\nu}_\mathrm{\ell}\gamma^\mu \nu_\mathrm{\ell}\right)\epsilon_\ell\;,
\label{epsilon_ell}
\end{eqnarray}
where we have dropped the subscript ``light" from the neutrino fields.

Lepton universality data
from $W$ decays and from charged current 
$\pi,\tau$ and $K$ decays \cite{Loinaz:2004qc} constraint
differences $\epsilon_{\ell_i}-\epsilon_{\ell_j}$. 
LEP/SLD and other precision electroweak 
data will imposed additional constraints on $\epsilon_\ell$ in combination with the oblique parameters,
as will NuSOnG.
A fit to all the charge current decay data and LEP/SLD with $S$, $T$,
$\epsilon_e$ and $\epsilon_\mu$ yields
\begin{eqnarray}
S & = & -0.05 \pm 0.11 \;,\cr
T & = & -0.44 \pm 0.28 \;,\cr
\epsilon_e & = & 0.0049 \pm 0.0022 \;,\cr
\epsilon_\mu & = & 0.0023 \pm 0.0021 \;.
\end{eqnarray}
%
%
%
%

If we now included hypothetical data from NuSOnG, assuming NuSOnG achieves its precision goals and measures central values consistent with the Standard Model, we see the constraints on $\epsilon_\mu$ and $\epsilon_e$ are substantially improved.  In this case, the fit yields
\begin{eqnarray}
S & = & \phantom{-}0.00 \pm 0.10 \;,\cr
T & = & -0.11 \pm 0.12 \;,\cr
\epsilon_e & = & 0.0030 \pm 0.0017 \;,\cr
\epsilon_\mu & = & 0.0001 \pm 0.0012. \;,
\end{eqnarray}
%

Fig.~\ref{epsilon_fit} shows the two dimensional cross section in
the $\epsilon_e$-$\epsilon_\mu$ plane of the four dimensional fit.
The likelihood coutours are 2D projections.   
Though not obvious from the figure, it is NuSOnG's improved measurement of $g_L^2$ which contributes the most to strengthening the bounds on the $\epsilon_\ell$.

In models of this class lepton
flavor violating decays such as $\mu \rightarrow e \gamma$ impose additional constraints on
products $\epsilon_{\ell_i} \epsilon_{\ell_j}$.
For example, the strong constraint from $\mu \rightarrow e \gamma$ implies $\epsilon_{e}
\epsilon_{\mu} \approx 0$.  This type of model has been proposed as a solution to the NuTeV anomaly.  If we take take only one of $\epsilon_e$ or $\epsilon_\mu$ to be nonzero (to respect the constraint from $\mu \rightarrow e \gamma$), the NuTeV value of $g_L^2$ is accommodated in the fit by best-fit values of $\epsilon$ that are large and positive and best-fit values of T are large and negative (consistent with a heavy Higgs).  

\subsection{Right-handed coupling of the neutrino to the $Z$}

In the Standard Model, neutrino
couplings to the $W$- and $Z$-bosons are purely left-handed.  The fact
that the neutrino coupling to the $W$-boson and an electron is purely
left-handed is, experimentally, a well-established fact (evidence
includes precision measurements of pion and muon decay, nuclear
processes, etc.). By contrast, the nature of the neutrino coupling to
the $Z$ boson is, experimentally, far from being precisely established
\cite{Carena:2003aj}. The possibility of a right-handed neutrino--$Z$-boson coupling is not included in the previous
discussions, and is pursued separately in this subsection.

\begin{figure*}
\centering
\scalebox{0.8}{\includegraphics[clip=true]{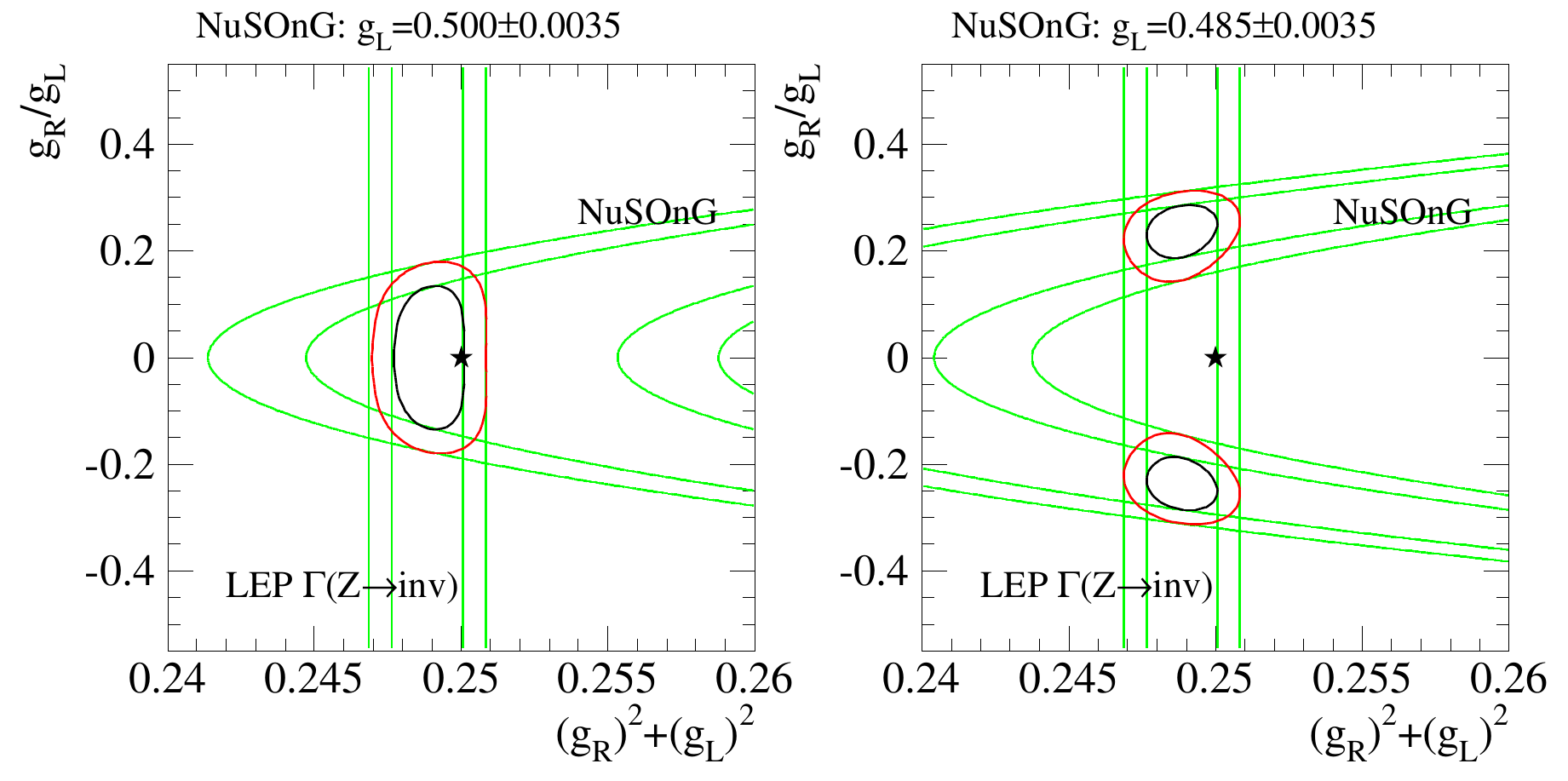}}
\vspace{1mm}
\caption{Precision with which the right-handed neutrino--$Z$-boson
  coupling can be determined by combining NuSOnG measurements of
  $g_L^{\nu}$ with the indirect determination of the invisible
  $Z$-boson width at LEP if (left) the $\nu+e$ scattering
  measurement is consistent with the Standard Model prediction
  $g_L^{\nu}=0.5$ and (right) the $\nu+e$
  scattering measurement is significantly lower, $g_L^{\nu}=0.485$,
  but still in agreement with the CHARM II measurement(at the one
  sigma level). Contours (black, red) are one and two sigma,
  respectively. The star indicates the Standard Model
  expectation.}
\label{glgr}
\end{figure*}

The best measurement of the neutrino coupling to the $Z$-boson is
provided by indirect measurements of the invisible $Z$-boson width at
LEP. In units where the Standard Model neutrino--$Z$-boson couplings are
$g_L^{\nu}=0.5$, $g_R^{\nu}\equiv 0$, the LEP measurement \cite{invZ} 
translates
into $(g^{\nu}_L)^2+(g^{\nu}_R)^2=0.2487\pm 0.0010$. Note that
this result places no meaningful bound on $g_R^{\nu}$.

Precise, model-independent information on $g^{\nu}_L$ can be obtained
by combining $\nu_{\mu}+e$ scattering data from CHARM II and LEP and
SLD data. Assuming model-independent couplings of the fermions to the
$Z$-boson, $\nu_{\mu} +e$ scattering measures $g_L^{\nu}=\sqrt{\rho}/2$, while
LEP and SLD measure the left and right-handed couplings of the
electron to the $Z$. The CHARM II result translates into
$|g_L^{\nu}|=0.502\pm 0.017$ \cite{Carena:2003aj}, assuming that the
charged-current weak interactions produce only left-handed
neutrinos. In spite of the good precision of the CHARM II result
(around 3.5\%), a combination of all available data allows
$|g_R^{\nu}/g_L^{\nu}|\sim 0.4$ at the two $\sigma$ confidence level
\cite{Carena:2003aj}.

Significant improvement in our understanding of $g_R^{\nu}$ can only
be obtained with more precise measurements of $\nu+e$ scattering, or
with the advent of a new high intensity $e^+e^-$ collider, such as the
ILC. By combining ILC running at the $Z$-boson pole mass and at
$\sqrt{s}=170$~GeV, $|g_R^{\nu}/g_L^{\nu}|\lesssim 0.3$ could be
constrained at the two $\sigma$ level after analyzing
$e^+e^-\to\gamma+$missing energy events \cite{Carena:2003aj}.

Assuming that $g^{\nu}_L$ can be measured with 0.7\% uncertainty, 
Fig.~\ref{glgr} depicts an estimate of how precisely $g_R^{\nu}$ could be constrained
once NuSOnG ``data'' is combined with LEP data. Fig.~\ref{glgr}(left) considers the hypothesis that the Standard Model expectations are correct. In this case, NuSOnG data would reveal that $g_R/g_L$ is less than 0.2 at the two sigma level. On the other hand, if $g_R/g_L=0.25$ -- in good agreement with the current CHARM II and LEP data -- NuSOnG data should reveal that $g_R\neq 0$ at more than the two sigma level, as depicted in Fig.~\ref{glgr}(right).

The capability of performing this measurement in other experiments has
been examined.  The NuSOnG measurement compares favorably, and
complements, the ILC capabilities estimated in \cite{Carena:2003aj}.
Ref~\cite{de Gouvea:2006cb} studied measurements using other neutrino
beams, including reactor fluxes and beta beams.  NuSOnG's reach is
equivalent to or exceeds the most optimistic estimates for these
various neutrino sources.

\section{Specific Theoretical Models and Experimental Scenarios \label{Models}}

If NuSOnG's measurements agree
with the SM within errors, we will place stringent constraints on new physics
models; if they disagree, it will be a signal for new physics.  In the
latter case the availability of both DIS and ES channels will improve
our ability to discriminate among new physics candidates.  NuSOnG will
also provide an important complement to the LHC.  The LHC will provide
detailed information about the spectrum of new states directly
produced.  However, measurements of the widths of these new states
will provide only limited information about their couplings.  NuSOnG
will probe in multiple ways the couplings of these new states to
neutrinos and to other SM particles.

In this section we provide several case studies of NuSOnG
sensitivity to specific models of new physics.  These include several
typical $Z^\prime$models, leptoquark models, models of R-parity
violating supersymmetry, and models with extended Higgs sectors.  We
examine how these will affect $\nu_\mu e$ ES and $\nu_\mu N$ DIS at
tree-level.  Our list is far from exhaustive but serves to illustrate
the possibilities.  We summarize our contributions in Table~\ref{Tab:modeloverview}.

\begin{center}
\begin{table*}[tbp] 
\begin{ruledtabular}
\begin{tabular}{|c|l|}
Model &   Contribution of NuSOnG Measurement \\ \hline \hline
Typical $Z^\prime$ Choices:  $(B-xL)$,$(q-xu)$,$(d+xu)$ &   At the level of, and complementary to, LEP II bounds.\\ \hline
Extended Higgs Sector & At the level of, and complementary to $\tau$ decay bounds. \\ \hline
R-parity Violating SUSY & Sensitivity to masses $\sim 2$ TeV at 95\% CL. \\
                        & Improves bounds on slepton couplings by $\sim 30\%$ and \\
                        & on some squark couplings by factors of 3-5. \\ \hline
Intergenerational Leptoquarks with non-degenerate masses & Accesses unique combinations of couplings.  \\ 
&  Also accesses coupling combinations explored by $\pi$ decay bounds, \\
&  at a similar level.\\ \hline
\end{tabular}
\end{ruledtabular}
\label{Tab:modeloverview}
\caption{Summary of NuSOnG's contribution in the case of specific models}
\end{table*}
\end{center}

The opposite way to approach this problem is to ask:  in the face
of evidence for new Terascale Physics, how can we differentiate 
between specific models?  NuSOnG has the  potential to {\em discover} new
physics through indirect probes, in the event that one or more of its
measurements definitively contradicts SM predictions.  We discuss
several possible patterns of deviation of model-independent parameters
from SM predictions and some interpretations in terms of particular
models.  This is presented in the context of various expectations 
for LHC to illustrate how NuSOnG enhances the overall physics program.
Since the NuTeV reanalysis is ongoing, and
since the ES constraints from CHARM-II are weak, it is prudent that we
commit to no strong assumptions about the central value of the NuSOnG
measurements but instead consider all reasonable outcomes.

\subsection{Sensitivity in the Case of Specific Theoretical Models}

\begin{figure}
\vspace{-0.75in}
\scalebox{0.45}{\includegraphics{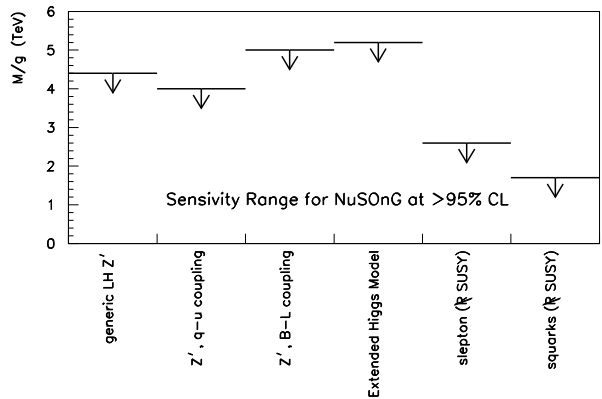}}
\vspace{-2.in}
\caption{\label{massreach} Some examples of NuSOnG's 2$\sigma$
  sensitivity to new high-mass particles commonly considered in the
  literature.  For explanation of these ranges, and further examples,
  see text.}
\end{figure}

We next consider the constraints imposed by the proposed NuSOnG
measurements on explicit models of BSM
physics.  An explicit model provides relations among effective
operators which give stronger and sometimes better-motivated
constraints on new physics than is obtained from bounds obtained by
considering effective operators one by one, but at the expense of the
generality of the conclusions.  Many models can be analyzed using the
effective Lagrangian of Eq.~(\ref{NSI}), but others introduce new
operators and must be treated individually.  The list of models
considered is not exhaustive, but rather illustrates the new physics
reach of NuSOnG.

\subsubsection{$Z^\prime$ models}

Massive $Z'$ fields are one of the simplest signatures of
physics beyond the Standard Model. (For a recent review, see
\cite{Langacker}.) $Z'$ vector bosons are generic in grand unified
theories and prevalent in theories that address the electroweak gauge
hierarchy.  They may stabilize the weak scale directly by canceling
off quadratic divergences of Standard Model fields, as in theories of
extra-dimensions or Little Higgs theories. In supersymmetric models,
$Z'$ fields are not needed to cancel quadratic divergences, but
are still often tied to the scale of soft-breaking (and hence the
electroweak scale). In these last two cases, the $Z'$ typically has a
TeV-scale mass, and is an attractive target for NuSOnG.

If the $Z'$ mass is sufficiently large, its exchange is well-described
at NuSOnG energies by the effective operator of
Eq.~(\ref{eq:leff}). In this case, the new physics scale is related to
the $Z'$ model by $\Lambda \sim M_{Z'}/g_{Z'}$, the ratio of the $Z'$
mass to its gauge-coupling. Further model-dependence shows up in the
ratio of fermion charges under the ${\rm U(1)'}$ symmetry associated
with the $Z'$, and the presence of any $Z-Z'$ mixing.  With reasonable
theoretical assumptions, the absence of new sources of large
flavor-changing neutral currents, the consistency of Yukawa
interactions, and anomaly cancellation with a minimal number of exotic
fermions, the number of interesting models can be reduced
substantially, to four discrete families of generic $U(1)'$ models
each containing one free parameter, $x$ \cite{CDDT}.  In
Table~\ref{Tab:zprime}, we indicate the charges of $\nu_{\mu L}, e_L,
e_R$ under these families of $U(1)'$ symmetries.

\begin{table}
\begin{ruledtabular}
\begin{tabular}{|c|c|c|c|c|}
&$U(1)_{B-xL}$ &$U(1)_{q+xu}$& $U(1)_{10+x\bar{5}}$& $U(1)_{d-xu}$ \\ \hline 
$\nu_{\mu L}, e_L$ & $-x$ & $-1$ & $x/3$ & $(-1+x)/3$ \\
$e_R$ & $-x$ & $-(2+x)/3$ & $-1/3$ & $x/3$ \\
\end{tabular}
\end{ruledtabular}
\label{Tab:zprime}
\caption{Charges of $\nu_{\mu L}, e_L, e_R$ under 4 phenomenologically viable classes of  $U(1)'$ symmetries.  Each value of $x$ corresponds to a different $U(1)'$ symmetry that is considered.}
\end{table}

Using the sensitivity of NuSOnG
to the scale $\Lambda$ in $\nu_{\mu}$ scattering shown in
Figure~\ref{fig:eff_limit}, we can bound the combination $M_{Z'}/g_{Z'}$ for the four families of $Z'$ models as a function of $x$. It is important to note that these bounds are competitive with the LEP-II bounds found in \cite{CDDT}, which are based on $Z'$ decays to all fermions, not just electrons and neutrinos.

\begin{figure}
\centering
\includegraphics[width=3.25in]{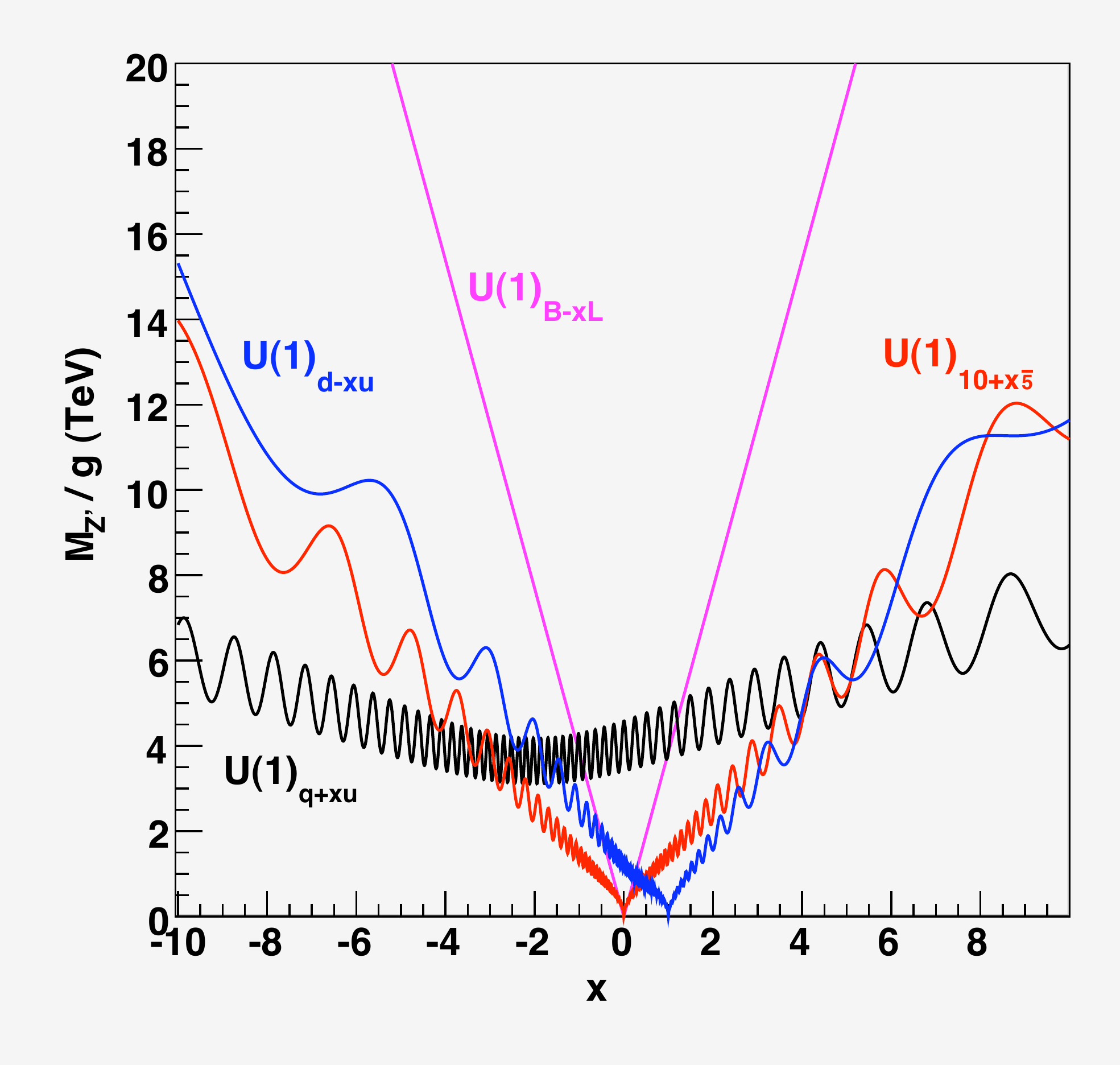}
\vspace{1mm}
\caption{95\% confidence level sensitivity of NuSOnG to the indicated
  $Z'$ models. The charges of the electrons and neutrinos under the
  underlying $U(1)'$ gauge symmetry are described in
  Table~\ref{Tab:zprime}.  The bounds are plotted as functions of the
  parameter $x$, which scans over allowed fermion charges for each
  family of $U(1)'$ symmetries, versus the ratio $M_{z'}/g_{Z'}$.  }
\label{fig:zprime}
\end{figure}

There are $Z^\prime$ models which distinguish among generations can
affect neutrino scattering.  These will be probed by NuSOnG at the TeV
scale~\cite{L1minusL2,LmuMinusLtau,Bminus3Ltau,Bminus3Le,Bminus3over2LtauplusLmu}.
Among these, $B-3L_\mu$ was suggested as a possible explanation for the
NuTeV anomaly~\cite{Ma:2001tb, Davidson:2001ji}, however, we show here
that this is not the case.  Nevertheless, it remains an interesting
example to consider.

In the gauged $B-3L_\mu$ the $Z'$ modifies $\nu_\mu N$ DIS. 
The exchange of the $Z'$ between the $\nu_\mu$ and the quarks
induces operators with coefficients
\begin{eqnarray}
\varepsilon_{\mu\mu}^{uL}\;&=&\;\varepsilon_{\mu\mu}^{uR}\;=\;
\varepsilon_{\mu\mu}^{dL}\;=\;\varepsilon_{\mu\mu}^{dR} \cr
&\;=\; &
-\frac{1}{2\sqrt{2}G_F}\frac{g_{Z'}^2}{M_{Z'}^2}
\;\equiv\;\varepsilon_{B-3L_\mu}\;.
\end{eqnarray}
which shift  $g_L^2$ and $g_R^2$ by
\begin{equation}
\Delta g_L^2=\Delta g_R^2=-\frac{2 s^2}{3}\,\varepsilon_{B-3L_\mu}.
\end{equation}
It should be noted that since $\varepsilon_{B-3L_\mu}$ is negative,
this shows that both $g_L^2$ and $g_R^2$ will be shifted positive.
This, in fact, excludes gauged $B-3L_\mu$ as an explanation of the
NuTeV anomaly.  
With this said, a NuSOnG measurement of $g_L^2$ and $g_R^2$ that
improves on NuTeV errors by a factor of 2 yields a $2\sigma$ bound
\begin{equation}
\frac{M_{Z'}}{g_{Z'}} \;>\; 2.2\,\mathrm{TeV}\;.
\end{equation}
which is comparable and complementary to the existing bound from D0,
and thus interesting to consider.

\subsubsection{Models with extended Higgs sectors}

In the Zee \cite{Zee:1980ai} and Babu-Zee \cite{Babu:1988ki} models,
an isosinglet scalar $h^+$ with hypercharge $Y=+1$ is introduced,
which couples to left-handed lepton doublets $\ell$ as
\begin{equation} \label{L_Babu}
\mathcal{L}_h
\;=\; \lambda_{ab} 
\left(\,\overline{\ell^c_{aL}}\,i\sigma_2\,\ell_{bL}^{\phantom{\mathrm{T}}}\,\right) h^+ + h.c.\;,
\end{equation}
where $(ab)$ are flavor indices: $a,b=e,\mu,\tau$. 
The exchange of a charged Higgs induces the effective operator from Eq.~(\ref{NSI}) which with coefficient
\begin{equation}
\varepsilon_{\mu\mu}^{eL}\;=\;-\frac{1}{\sqrt{2}G_F}\frac{|\lambda_{e\mu}|^2}{M_h^2}\;,\qquad
\varepsilon_{\mu\mu}^{eR}\;=\; 0\;.
\end{equation}
From Eq.~(\ref{EPSePbounds}), the 95\% bound is:
\begin{equation}
\frac{M_h}{|\lambda_{e\mu}|} \;>\; 5.2\,\mathrm{TeV},\;.
\end{equation}
competitive with current bound from $\tau$-decay of $5.4$~TeV.

\subsubsection{R-parity violating SUSY}

Assuming the particle content of the 
Minimal Supersymmetric Standard Model (MSSM),
the most general R-parity violating superpotential (involving only
tri-linear couplings) has the form
\cite{Rparity_notations}
\begin{equation}
W_{\not R}
=\frac{1}{2}\lambda_{ijk}\hat{L}_i \hat{L}_j \hat{E}_k
+\lambda^{\prime}_{ijk}\hat{L}_i \hat{Q}_j \hat{D}_k
+\frac{1}{2}\lambda^{\prime\prime}_{ijk}\hat{U}_i \hat{D}_j \hat{D}_k\;,
\label{RPVlagrangian}
\end{equation}
where $\hat{L}_i$, $\hat{E}_i$, $\hat{Q}_i$, $\hat{D}_i$, and $\hat{U}_i$ are
the left-handed MSSM superfields defined in the usual fashion,
and the subscripts $i,j,k=1,2,3$ are the generation indices.
$SU(2)_L$ gauge invariance requires
the couplings $\lambda_{ijk}$ to be antisymmetric in the first two indices:
\begin{equation}
\lambda_{ijk} \;=\; -\lambda_{jik}\;,
\end{equation}
The purely baryonic operator $\hat{U}_i\hat{D}_j\hat{D}_k$ is irrelevant to neutrino scattering, so only the 
9 $\lambda_{ijk}$ and 27 $\lambda'_{ijk}$ couplings are of interest.

From the $\hat{L}\hat{L}\hat{E}$ part of the Eq.~(\ref{RPVlagrangian})
slepton exchange will contribute to $\nu_\mu e$ ES at NuSOnG.
These induce four-fermion operators appearing in Eq.~(\ref{NSI}) with corresponding coefficients
\begin{eqnarray}
\varepsilon_{\mu\mu}^{eL} & \;=\; &
-\frac{1}{4\sqrt{2}G_F}\sum_{k=1}^3 \frac{|\lambda_{21k}|^2}{M_{\tilde{e}_{kR}}^2}\;,\qquad \cr
\varepsilon_{\mu\mu}^{eR} & \;=\; & 
+\frac{1}{4\sqrt{2}G_F}\sum_{j=1,3} \frac{|\lambda_{2j1}|^2}{M_{\tilde{e}_{jL}}^2}\;.
\end{eqnarray}
If we place bounds on the sleptons one at a time, then
Eq.~(\ref{EPSePbounds}) translates to the $2 \sigma$ bounds shown in
Table~\ref{Tab:sleptons}, presented for masses of 100 GeV.  To rescale
to different masses, use $\left(\frac{M}{100\,\mathrm{GeV}}\right) $.
This can be compared to current bounds Ref.~\cite{Barbier:2004ez}.
NuSOnG improves all of these bounds.

\begin{table}
\begin{tabular}{|c||c|l|}
\hline
\ Coupling\ \ &\ 95\% NuSOnG bound\ \ & current 95\% bound\ \ \\
\hline\hline
$|\lambda_{121}|$ & $0.03$ & \ $0.05$ ($V_{ud}$) \\
$|\lambda_{122}|$ & $0.04$ & \ $0.05$ ($V_{ud}$) \\
$|\lambda_{123}|$ & $0.04$ & \ $0.05$ ($V_{ud}$) \\
$|\lambda_{231}|$ & $0.05$ & \ $0.07$ ($\tau$ decay) \\
\hline\hline
$|\lambda'_{211}|$ & $0.05$ & \ $0.06$ ($\pi$ decay) \\
$|\lambda'_{212}|$ & $0.06$ & \ $0.06$ ($\pi$ decay) \\
$|\lambda'_{213}|$ & $0.06$ & \ $0.06$ ($\pi$ decay) \\
$|\lambda'_{221}|$ & $0.07$ & \ $0.21$ ($D$ meson decay)\ \ \\
$|\lambda'_{231}|$ & $0.07$ & \ $0.45$ ($Z\rightarrow \mu^+\mu^-$)\\
\hline
\end{tabular}
\caption{Potential bounds on the R-parity violating $LLE$ (top) and  $LQD$ (bottom) couplings
from NuSOnG, assuming that only one coupling is non-zero at a time  for each set.
All squark and slepton masses are set to 100~GeV.  To obtain limits for
different masses, rescale by $\left(\frac{M}{100\,\mathrm{GeV}}\right) $.  Current bounds are from Ref.~\cite{Barbier:2004ez}.}
\label{Tab:sleptons}
\end{table}

From the $\hat{L}\hat{Q}\hat{D}$ part of Eq.~(\ref{RPVlagrangian}), squark exchange will
contribute to contribute to NC $\nu_\mu N$  DIS and  CC $\nu_\mu N$ DIS.  
The resulting shifts in $g_L^2$ and $g_R^2$ are
\begin{eqnarray}
\delta g_L^2 & = & 
2\Bigl[\, g_L^{\nu d} \varepsilon_{\mu\mu}^{dL} - g_L^2 \varepsilon_c \,\Bigr] \;,\cr
\delta g_R^2 & = &
2\Bigl[\, g_R^{\nu d} \varepsilon_{\mu\mu}^{dR} - g_R^2 \varepsilon_c \,\Bigr]\;,
\end{eqnarray}
where
\begin{eqnarray}
\varepsilon_{\mu\mu}^{dL} & = &
-\frac{1}{4\sqrt{2}G_F}\sum_{k=1}^3 \frac{|\lambda'_{21k}|^2}{M_{\tilde{d}_{kR}}^2}\;,\cr
\varepsilon_{\mu\mu}^{dR} & = &
-\frac{1}{4\sqrt{2}G_F}\sum_{j=1}^3 \frac{|\lambda'_{2j1}|^2}{M_{\tilde{d}_{jL}}^2}
\;,\cr
\varepsilon_c & = &
+\frac{1}{4\sqrt{2}G_F}\sum_{k=1}^3 \frac{|\lambda'_{21k}|^2}{M_{\tilde{d}_{kR}}^2}
\;=\; -\varepsilon_{\mu\mu}^{dL}\;,
\end{eqnarray}
$\varepsilon_{\mu\mu}^{dL}$ and $\varepsilon_{\mu\mu}^{dR}$ are
associated with terms of Eq.~(\ref{NSI}), while $\varepsilon_c$ is
associated with a four-fermion interaction that corrects charged
currents,
\begin{equation}
-2\sqrt{2}G_F\varepsilon_c 
\Bigl[
\bigl( \overline{\mu_L} \gamma_\sigma \nu_{\mu L} \bigr)
\bigl( \overline{u_L}\gamma^\sigma d_L \bigr) 
+ h.c. \;
\Bigr] \;.
\end{equation}
The shifts in $g_L^2$ and $g_R^2$ are:
\begin{eqnarray}
\delta g_L^2 & = & 
2\left(g_L^{\nu d}+g_L^2\right) \varepsilon_{\mu\mu}^{dL} \;,\cr
\delta g_R^2 & = &
2g_R^2 \varepsilon_{\mu\mu}^{dL} 
+ 2g_R^{\nu d} \varepsilon_{\mu\mu}^{dR}
\;.
\end{eqnarray}
Assuming the projected precision goals for NuSOnG on $g_L^2$ and
$g_R^2$, and allowing only one of the couplings to be nonozero at a
time, the $2 \sigma$ bounds are given in Table~\ref{Tab:sleptons} mass
of 100~GeV, in all cases.  To obtain limits for different masses,
one simply rescales by $\left(\frac{M}{100\,\mathrm{GeV}}\right) $.
NuSOnG's measurements are competitive with $\pi$ decay bounds, and 
improves the current bounds on the $221$ and $231$ couplings by factors
of 3 and 5, respectively.

\subsubsection{Intergenerational leptoquark models}

Measurements of $g_L^2$ and $g_R^2$ are sensitive to leptoquarks.
Because the exchange of a leptoquark can interfere with both $W$ and
$Z$ exchange processes, we cannot use the limits on the NSI's of
Eq.~(\ref{NSI}), since we must also include the effects of 
the four-fermion operators associated with charged-current processes.
Instead, the interactions of leptoquarks with
ordinary matter can be described in a model-independent fashion by an
effective low-energy Lagrangian as discussed in
Refs.~\cite{BuchRuckWyler,leptoquarks} for generation-universal leptoquark couplings.  
For leptoquarks to contribute to $\nu_\mu N$
DIS, they must couple second generation leptons to first generation
quarks, so we use the more general Lagrangian of
~\cite{Davidson:1993qk,Honda:2007wv},
which allows the coupling constants to depend
on the generations of the quarks and leptons that couple to each
leptoquark.  
We summarize the quantum numbers and couplings of the various leptoquarks fields 
in Table~\ref{LQtable};
our notation conventions are those of Ref.~\cite{Honda:2007wv}.

\begin{center}
\begin{table*}[tbp]
\begin{ruledtabular}
\begin{tabular}{|c|c||c|c|c|c|c||c||c|}
\hline
\multicolumn{2}{|c|}{\ Leptoquark\ \ }  &\ Spin\ \ & $\;\;F\;\;$ & 
$\,SU(3)_C\,$
& $\quad I_3\quad$ & $\quad Y\quad$ & $\;\;Q_{em}\;\;$ &\ Allowed 
Couplings\ \\
\hline\hline
$\;S_1\;$
& $\;S_1^0\;$
& $0$
& $-2$
& $\bar{3}$
& $\phantom{+}0$
& $\phantom{+}\frac{1}{3}$
& $\phantom{+}\frac{1}{3}$
& 
$\;g_{1L}(\overline{u_L^c}e_L^{\phantom{c}}-\overline{d_L^c}\nu_L^{\phantom{c}}),\,
g_{1R}(\overline{u^c_R}e_R^{\phantom{c}})\;$ \\
\hline
$\;\tilde{S}_1\;$
& $\;\tilde{S}_1^0\;$
& $0$
& $-2$
& $\bar{3}$
& $\phantom{+}0$
& $\phantom{+}\frac{4}{3}$
& $\phantom{+}\frac{4}{3}$
& $\tilde{g}_{1R}(\overline{d_R^c}e_R^{\phantom{c}})$ \\
\hline
$\;V_{2\mu}\;$
& $\;V_{2\mu}^+\;$
& $1$
& $-2$
& $\bar{3}$
& $+\frac{1}{2}$
& $\phantom{+}\frac{5}{6}$
& $\phantom{+}\frac{4}{3}$
& $g_{2L}(\overline{d_R^c}\gamma^\mu e_L^{\phantom{c}}),
\,g_{2R}(\overline{d_L^c}\gamma^\mu e_R^{\phantom{c}}) $ \\
& $\;V_{2\mu}^-\;$
&
&
&
& $-\frac{1}{2}$
&
& $\phantom{+}\frac{1}{3}$
& $g_{2L}(\overline{d_R^c}\gamma^\mu \nu_L^{\phantom{c}}),\,
g_{2R}(\overline{u_L^c}\gamma^\mu e_R^{\phantom{c}}) $ \\
\hline
$\tilde{V}_{2\mu}\;$
& $\tilde{V}_{2\mu}^+\;$
& $1$
& $-2$
& $\bar{3}$
& $+\frac{1}{2}$
& $-\frac{1}{6}$
& $\phantom{+}\frac{1}{3}$
&  $\tilde{g}_{2L}(\overline{u_R^c}\gamma^\mu e_L^{\phantom{c}})$\\
& $\;\tilde{V}_{2\mu}^-\;$
&
&
&
& $-\frac{1}{2}$
&
& $-\frac{2}{3}$
& $\tilde{g}_{2L}(\overline{u_R^c}\gamma^\mu \nu_L^{\phantom{c}})$  \\
\hline
$\;\vec{S}_3\;$
& $\;S_3^+\;$
& $0$
& $-2$
& $\bar{3}$
& $+1$
& $\phantom{+}\frac{1}{3}$
& $\phantom{+}\frac{4}{3}$
& $-\sqrt{2} g_{3L}(\overline{d_L^c}e_L^{\phantom{c}}) $ \\
& $\;S_3^0\;$
&
&
&
& $\phantom{+}0$
&
& $\phantom{+}\frac{1}{3}$
& 
$-g_{3L}(\overline{u_L^c}e_L^{\phantom{c}}+\overline{d_L^c}\nu_L^{\phantom{c}}) 
$ \\
& $\;S_3^-\;$
&
&
&
& $-1$
&
& $-\frac{2}{3}$
& $\sqrt{2}g_{3L}(\overline{u_L^c}\nu_L^{\phantom{c}})$ \\
\hline\hline
$\;S_2\;$
& $\;S_2^+\;$
& $0$
& $0$
& $3$
& $+\frac{1}{2}$
& $\phantom{+}\frac{7}{6}$
& $\phantom{+}\frac{5}{3}$
&  $h_{2L}(\overline{u_R} e_L),h_{2R}(\overline{u_L}e_R) $ \\
& $\;S_2^-\;$
&
&
&
& $-\frac{1}{2}$
&
& $\phantom{+}\frac{2}{3}$
& $h_{2L}(\overline{u_R} \nu_L),-h_{2R}(\overline{d_L}e_R) $ \\
\hline
$\;\tilde{S}_2\;$
& $\;\tilde{S}_2^+\;$
& $0$
& $0$
& $3$
& $+\frac{1}{2}$
& $\phantom{+}\frac{1}{6}$
& $\phantom{+}\frac{2}{3}$
& $\tilde{h}_{2L}(\overline{d_R}e_L)$ \\
& $\;\tilde{S}_2^-\;$
&
&
&
& $-\frac{1}{2}$
&
& $-\frac{1}{3}$
& $\tilde{h}_{2L}(\overline{d_R}\nu_L)$ \\
\hline
$\;V_{1\mu}\;$
& $\;V_{1\mu}^0\;$
& $1$
& $0$
& $3$
& $\phantom{+}0$
& $\phantom{+}\frac{2}{3}$
& $\phantom{+}\frac{2}{3}$
& $\;h_{1L}(\overline{u_L}\gamma^\mu \nu_L+\overline{d_L}\gamma^\mu 
e_L),\;h_{1R}(\overline{d_R}\gamma^\mu e_R)\;$ \\
\hline
$\;\tilde{V}_{1\mu}\;$
& $\;\tilde{V}_{1\mu}^0$
& $1$
& $0$
& $3$
& $\phantom{+}0$
& $\phantom{+}\frac{5}{3}$
& $\phantom{+}\frac{5}{3}$
& $\tilde{h}_{1R}(\overline{u_R}\gamma^\mu e_R)$ \\
\hline
$\;\vec{V}_{3\mu}\;$
& $\;V_{3\mu}^+\;$
& $1$
& $0$
& $3$
& $+1$
& $\phantom{+}\frac{2}{3}$
& $\phantom{+}\frac{5}{3}$
& $\sqrt{2}h_{3L}(\overline{u_L}\gamma^\mu e_L)$ \\
& $\;V_{3\mu}^0\;$
&
&
&
& $\phantom{+}0$
& 
& $\phantom{+}\frac{2}{3}$
& $h_{3L}(\overline{u_L}\gamma^\mu \nu_L-\overline{d_L}\gamma^\mu e_L) $ \\
& $V_{3\mu}^-\;$
&
&
&
& $-1$
& 
& $-\frac{1}{3}$
& $\sqrt{2}h_{3L}(\overline{d_L}\gamma^\mu \nu_L)$ \\
\hline
\end{tabular}
\caption[]{Quantum numbers of scalar and vector leptoquarks with
$SU(3)_C\times SU(2)_L\times U(1)_Y$ invariant couplings to quark-lepton
pairs ($Q_{\rm em}=I_3+Y$) \cite{PDG}.}
\label{LQtable}
\end{ruledtabular}
\end{table*}
\end{center}

\begin{figure}[ht]
\includegraphics[scale=0.75]{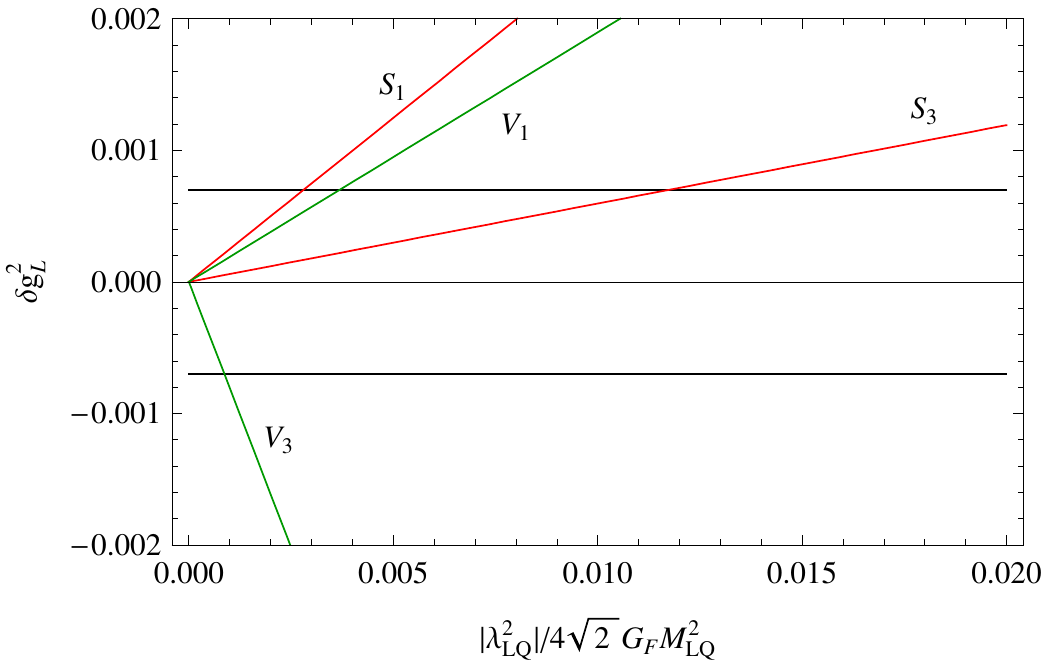}
\includegraphics[scale=0.75]{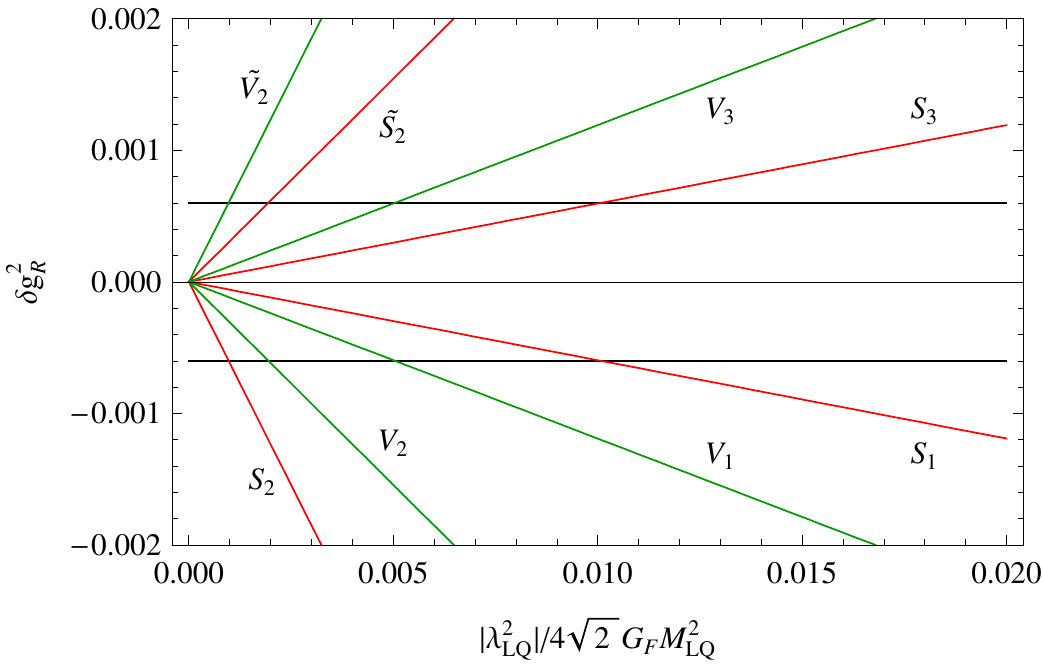}
\caption{Shifts in $g_L^2$ and $g_R^2$ due to leptoquarks.
Horizontal lines indicate the projected $1\sigma$ limits of NuSOnG.}
\label{gL2gR2shifts}
\end{figure}

The four-fermion operators induced by leptoquark exchange
will affect NC and/or CC processes, and at NuSOnG the effect manifests 
itself in shifts $g_L^2$ and $g_R^2$.
Assuming degenerate masses within each iso-multiplet,
the shifts in $g_L^2$ and $g_R^2$ can be written generically as
\begin{eqnarray}
\delta g_L^2 & = & C_L\,\frac{|\lambda_{LQ}^{12}|^2/M_{LQ}^2}{g^2/M_W^2}
\;=\; \frac{C_L}{4\sqrt{2}G_F}\frac{|\lambda_{LQ}^{12}|^2}{M_{LQ}^2} \;,\cr
\delta g_R^2 & = & C_R\,\frac{|\lambda_{LQ}^{12}|^2/M_{LQ}^2}{g^2/M_W^2}
\;=\; \frac{C_R}{4\sqrt{2}G_F}\frac{|\lambda_{LQ}^{12}|^2}{M_{LQ}^2} \;,
\end{eqnarray}
where $\lambda_{LQ}^{12}$ denotes the $(ij)=(12)$ coupling of the leptoquark
and $M_{LQ}$ is its mass.
In table~\ref{LQbounds} we list what they are, and in 
figure~\ref{gL2gR2shifts}
we plot the dependence of $\delta g_L^2$ and $\delta g_R^2$ on
the ratio $|\lambda_{LQ}|^2/M_{LQ}^2$.
Table~\ref{LQbounds} also lists the projected NuSOnG bounds on the 
coupling constants~\cite{LoinazProninTakeuchiprep}.
Existing bounds on $S_1$, $\vec{S}_3$, $V_1$, and $\vec{V}_3$ couplings 
from
$R_\pi = Br(\pi\rightarrow e\nu)/Br(\pi\rightarrow \mu\nu)$ are already much
stronger, but could be circumvented for $\vec{S}_3$ and $\vec{V}_3$
if the masses within the multiplet are allowed to be non-degenerate.

\begin{center}
\begin{table*}[ht]
\begin{ruledtabular}
\begin{tabular}{|c||c|c||c||c|c|}
\hline
$\;LQ\;$ & $\qquad C_L\qquad$ & $\quad C_R\quad$ & 
$\;\;|\lambda_{LQ}|^2\;\;$
&\ NuSOnG 95\% bound\ \
&\ 95\% bound from $R_\pi$ \ \ \\
\hline\hline
$S_1$
& $\;s^2\left(\frac{4}{3}-\frac{10}{9}s^2\right)\;$
& $-\frac{10}{9}s^4$
& $|g_{1L}^{12}|^2$
& $0.0036$
& $0.0037$ \\
\hline
$\vec{S}_3$
& $+\frac{10}{9}s^4$
& $+\frac{10}{9}s^4$
& $|g_{3L}^{12}|^2$
& $0.010$
& $0.0008$ \\
\hline
$S_2$
& $0$
& $-\frac{8}{3}s^2$
& $|h_{2L}^{12}|^2$
& $0.0013$
& N/A \\
\hline
$\tilde{S}_2$
& $0$
& $+\frac{4}{3}s^2$
& $|\tilde{h}_{2L}^{12}|^2$
& $0.0026$
& N/A \\
\hline
$V_1$
& $\;s^2\left(\frac{4}{3}-\frac{20}{9}s^2\right)\;$
& $-\frac{20}{9}s^4$
& $|h_{1L}^{12}|^2$
& $0.0040$
& $0.0018$ \\
\hline
$\vec{V}_3$
& $\;-4s^2\left(1-\frac{5}{9}s^2\right)\;$
& $+\frac{20}{9}s^4$
& $|h_{3L}^{12}|^2$
& $0.0011$
& $0.0004$ \\
\hline
$V_2$
& $0$ 
& $-\frac{4}{3}s^2$
& $|g_{2L}^{12}|^2$
& $0.0026$
& N/A \\
\hline
$\tilde{V}_2$
& $0$
& $+\frac{8}{3}s^2$
& $|\tilde{g}_{2L}^{12}|^2$
& $0.0013$
& N/A \\
\hline
\end{tabular}
\caption{Potential and existing 95\% bounds on the leptoquark couplings 
squared
when the leptoquark masses are set to 100~GeV.  To obtain the limits for
different leptoquark masses, multiply by $(M_{LQ}/100\,\mathrm{GeV})^2$.
Existing bounds on the $S_1$, $\vec{S}_3$, $V_1$, and $\vec{V}_3$ 
couplings from
$R_\pi=Br(\pi\rightarrow e\nu)/Br(\pi\rightarrow \mu\nu)$ are also shown.}
\label{LQbounds}
\end{ruledtabular}
\end{table*}
\end{center}

\subsection{Interplay with LHC to Isolate the Source of New Physics}

By the time NuSOnG runs, the LHC will have accumulated a wealth of
data and will have begun to change the particle physics landscape.
The message from LHC data may be difficult to decipher, however.  As
discussed below, NuSOnG will be able to help elucidate the new physics
revealed at the LHC.  The discovery of a Higgs along with the
anticipated measurement of the top mass to 1 GeV precision would
effectively fix the center of the $ST$ plot and will enhance the power
of the precision electroweak data as a tool for discovering new
physics.  If additional resonances are discovered at the LHC, it is
still likely that little will be learned about their couplings.

The NuSOnG experiment provides complementary information to LHC.
Rather than generalize, to illustrate the power of NuSOnG, two
specific examples are given here.  We emphasize that these are just
two of a wide range of examples, but they serve well to demonstrate
the point.    Here we have chosen examples from typical new physics
models other than $Z^\prime$ models which were discussed above, in
order to demonstrate the physics range which can be probed
by NuSOnG.

\begin{figure}[t]
\centering
{\includegraphics[width=3.25in, bb=100 100 650 500]{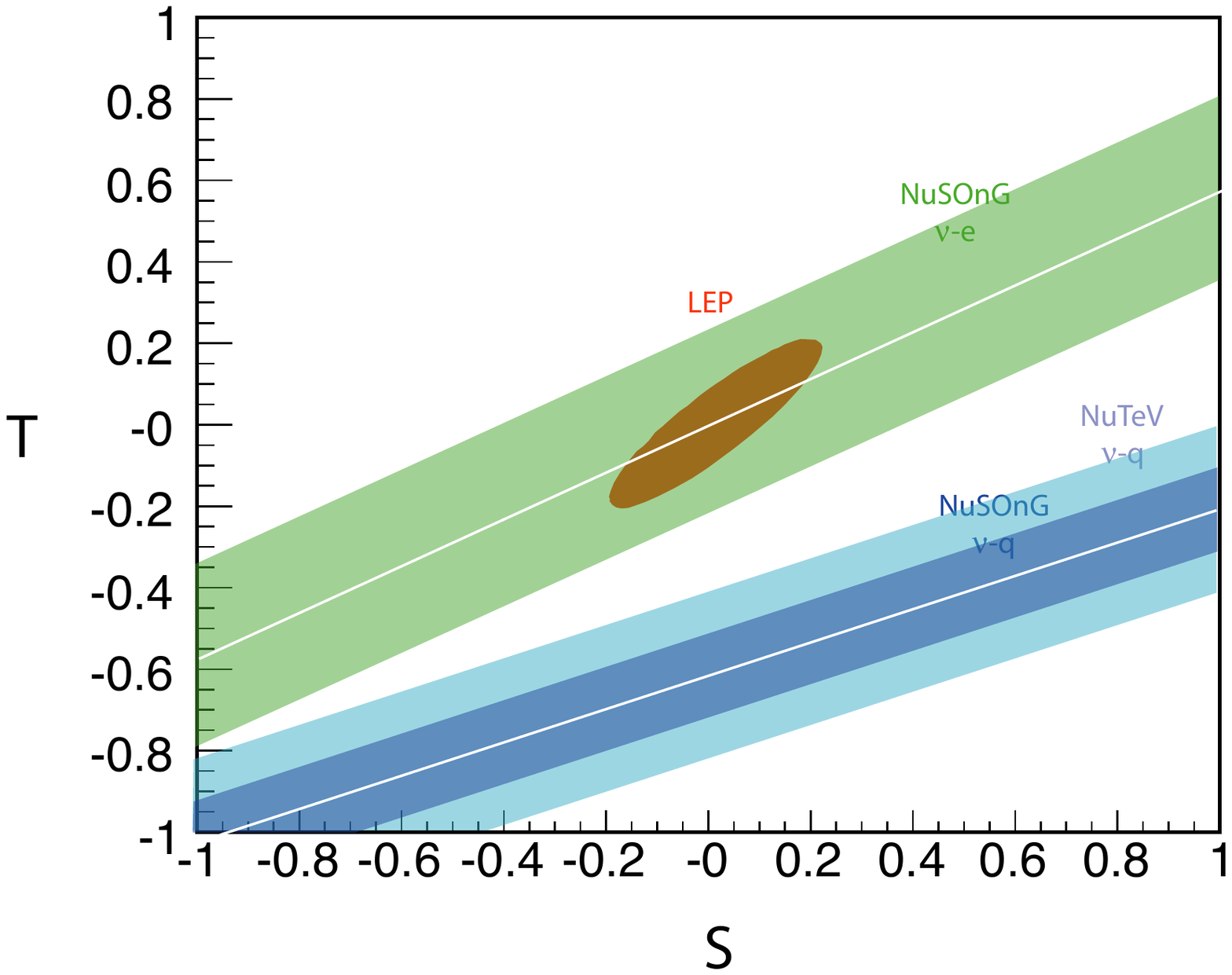}}
\caption{NuSOnG expectation in the case of a Tev-scale triplet
  leptoquark.  For clarity, this plot and the two following cases, show
  the expectation from only the two highest precision measurements
  from NuSOnG: $g_L^2$ and $\nu$ ES.}
\label{fig:NuSOnGEW1}
\end{figure}

First, extend the Standard Model to include a non-degenerate $SU(2)_L$
triplet leptoquark ($\vec{S}_3$ or $\vec{V}_3$ in the notation of
\cite{BuchRuckWyler}, with masses in the 0.5-1.5 TeV range.  At the
LHC these leptoquarks will be produced primarily in pairs through
gluon fusion, and each leptoquark will decay to a lepton and a jet
\cite{Belyaev}.  The peak in the lepton-jet invariant mass
distribution will be easily detected over background.  This will
provide the leptoquark masses but yield little information about their
couplings to fermions.  The leptoquarks will also shift the
neutrino-nucleon effective coupling $g_L^2$ in a way that depends
sensitively on both the leptoquark couplings and masses.  Such a
leptoquark-induced shift could provide an explanation for the NuTeV
anomaly \cite{Davidson:1993qk,Gabrielli:2000te,Davidson:2001ji}.  In this scenario, NuSOnG would find that
isospin and the strange sea can be constrained to the point that they
do not provide an explanation for the NuTeV anomaly, thus the NuTeV anomaly
is the
result of new physics. The NuSOnG PW measurement of $\sin2{\theta_W}$
will agree with NuTeV; $g_R^2$ and the $\nu e$ and $\overline{\nu}e$
elastic scattering measurements will agree with LEP.
Fig.~\ref{fig:NuSOnGEW1} illustrates this example.  NuSOnG's
measurement of $g_L^2$ would provide a sensitive measurement of the
leptoquark couplings when combined with the LHC mass measurements as
inputs.

\begin{figure}[t]
\centering
{\includegraphics[width=3.25in, bb=100 100 650 500]{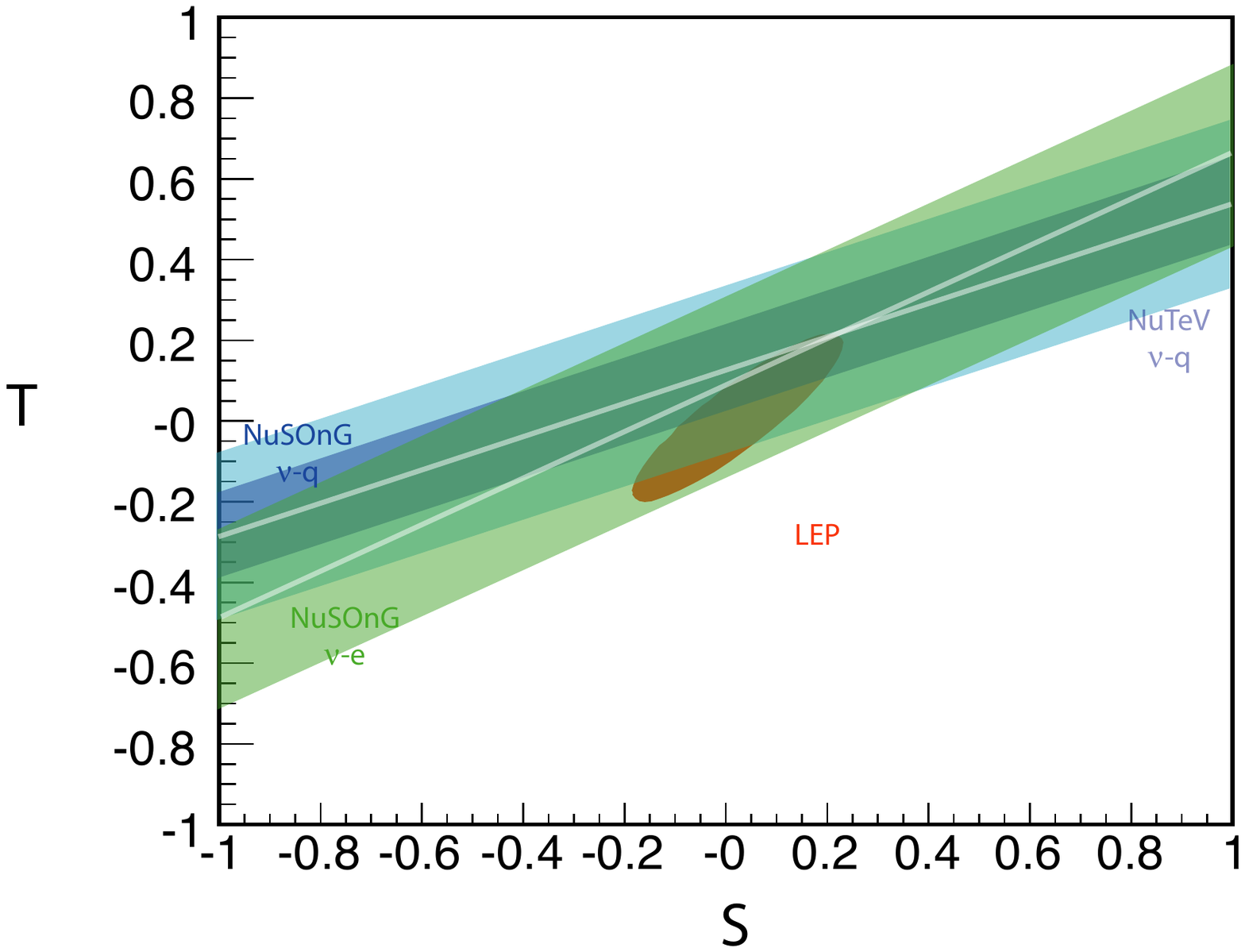}}
\caption{NuSOnG expectation if the NuTeV anomaly is
  due to isospin violation and
  there is a heavy 4th generation with isospin violation.}
\label{fig:NuSOnGEW2}
\end{figure}

\begin{figure}[t]
\centering
{\includegraphics[width=3.25in, bb=100 100 650 500]{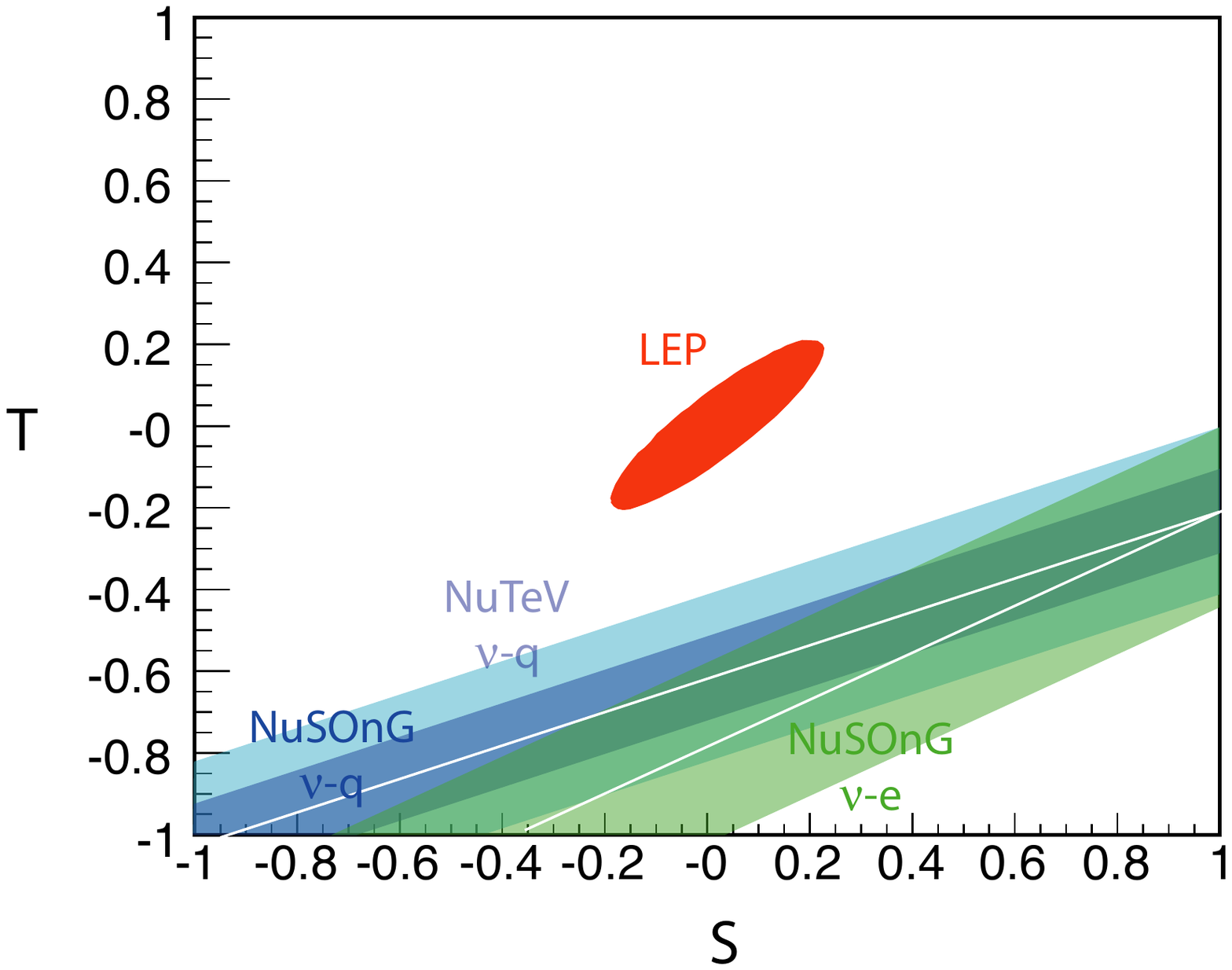}}
\caption{If LHC sees a Standard Model Higgs and no evidence of 
new physics, NuSOnG may reveal new physics in the neutrino sector.}
\label{fig:NuSOnGEW3}
\end{figure}

A second example is the existence of a fourth generation family.  A
fourth family with non-degenerate masses ({\it i.e.} isospin
violating) is allowed within the LEP/SLD constraints \cite{Tait}.  As
a model, we choose a fourth family with mass splitting on the order of
$\sim 75$ GeV and a 300 GeV Higgs. This is consistent with LEP at
1$\sigma$ and perfectly consistent with $M_W$, describing the point
(0.2,0.19) on the $ST$ plot.  In this scenario, LHC will
measure the Higgs mass from the highly enhanced $H \rightarrow ZZ$
decay. An array of exotic decays which will be difficult
to fully reconstruct, such as production of 6 W's and 2 b's, will be
observed at low rates.  In this scenario,
isospin violation explains the NuTeV anomaly, thus
the NuTeV PW  and the NuSOnG PW measurements agree with the
$\nu$eES measurements. These three precision neutrino results, all
with ``LEP-size'' errors, can be combined and will intersect the
one-sigma edge of the LEP measurements.
Fig.~\ref{fig:NuSOnGEW2} illustrates this example.  From this, the
source, a fourth generation with isospin violation, can be
demonstrated.

Lastly, while it seems unlikely, it is possible that LHC will observe
a Standard Model Higgs and no signatures of new physics.  If this is
the case, it is still possible for NuSOnG to add valuable clues to new
physics.  This is because the experiment is uniquely sensitive to the
neutrino sector.  If a situation such as is illustrated on
Fig.~\ref{fig:NuSOnGEW3} arose, the only explanation would be new
physics unique to neutrino interactions.


\section{Summary and Conclusions}

NuSOnG is an experiment which can search for new physics from keV
through TeV energy scales, as well as make interesting QCD
measurements.  This article has focussed mainly on the Terascale
physics which can be accessed through this new high energy, high
statistics neutrino scattering experiment.  The case has been made
that this new neutrino experiment would be a valuable addition to the
presently planned suite of experiments with Terascale reach.

The NuSOnG experiment design draws on the heritage of the CHARM II and
CCFR/NuTeV experiments.  A high energy, flavor-pure neutrino flux is
produced using 800 GeV protons from the Tevatron. The detector
consists of four modules, each composed of a finely-segmented
glass-target (SiO$_2$) calorimeter followed by a muon spectrometer.
In its five-year data acquisition period, this experiment will record
almost one hundred thousand neutrino-electron elastic scatters and
hundreds of millions of deep inelastic scattering events, exceeding the
current world data sample by more than an order of magnitude.  This
experiment can address concerns related to model systematics of
electroweak measurements in neutrino-quark scattering by direct
constraints using {\it in-situ} structure function measurements.

NuSOnG will be unique among present and planned experiments for its
ability to probe neutrino couplings to Beyond Standard Model
physics.  This experiment offers four distinct and complementary
probes of $S$ and $T$.  Two are of high precision with the proposed
run-plan, and the precision of the other two would be improved by a
follow-up five-year antineutrino run.  Neutrino-lepton non-standard
interactions can be probed with an order of magnitude improvement in
the measured effective couplings.  Neutrino-quark non-standard
interactions can be probed by an improvement in the measured
neutrino-quark effective couplings of a factor of two or better.  The
experiment is sensitive to new physics up to energy scales $\sim 5$
TeV at 95\% CL.  The measurements are sensitive to universality of the
couplings and an improvement in the $e$-family of 30\% and
$\mu$-family of 75\% will allow for probes of neutrissimos. As a
unique contribution, NuSOnG measures $g_R/g_L$, which is not
accessible by other near-future experiments.  This article described
NuSOnG's physics contribution under several specific models.  These
included models of $Z^\prime$s, extended Higgs models, leptoquark
models and $R$-parity violating SUSY models.  We also considered how,
once data are taken at LHC and NuSOnG, the underlying physics can be
extracted.  The opportunity for direct searches related to these
indirect electroweak searches was also described.  The conclusion of
our analysis is that a new neutrino experiment, such as NuSOnG, would
substantially enhance the presently planned Terascale program.

\bigskip

\begin{acknowledgments}
  We thank the following people for their thoughtful comments on
  the development of this physics case: 
  P. Langacker, M.~Shaposhnikov, F. Vannucci, J. Wells.

We acknowledge the support of the following funding agencies for the 
authors of this paper:  Deutsche Forschungsgemeinschaft, The Kavli Institute
for Theoretical Physics, The United States Department of Energy,  
The United States National Science Foundation.

\end{acknowledgments}


\end{document}